\documentclass[11pt,a4paper]{article}%
\usepackage{amssymb,amsmath, amsfonts}
\usepackage{graphicx,graphics}
\usepackage{mathtools}
\usepackage{bbold}
\usepackage[english]{babel}
\usepackage[utf8]{inputenc}
\usepackage{epsfig,url}
\usepackage{bbm,theorem}
\usepackage{a4wide}
\usepackage{color}
\usepackage{enumerate}
\usepackage{calrsfs}
\usepackage[bookmarks=true]{hyperref}
\usepackage{bookmark}
\usepackage{amsmath}
\usepackage{amsfonts}
\usepackage{amssymb}
\usepackage{verbatim}
\usepackage{graphicx}%
\providecommand{\U}[1]{\protect\rule{.1in}{.1in}}
\DeclareMathAlphabet{\pazocal}{OMS}{zplm}{m}{n}
\newtheorem{theorem}{Theorem}[section]
\newtheorem{conjecture}{Conjecture}[section]
\newtheorem{definition}[theorem]{Definition}

{\theorembodyfont{\upshape}
\newtheorem{remark}[theorem]{Remark}

}
\numberwithin{equation}{section}
\numberwithin{theorem}{section}
\newcommand{\qed}{\hfill$\Box$}

\newenvironment{proofof}[1][Proof]{\noindent \textit{#1.} }{\ \qed}

\newcommand{\R}{{\mathbb R}}

\DeclareMathOperator{\Tr}{Tr}

\newcommand{\beq}{\begin{equation}}
\newcommand{\eeq}{\end{equation}}
\newcommand{\beqs}{\begin{eqnarray}}
\newcommand{\eeqs}{\end{eqnarray}}

\newcommand{\calH}{{\cal H}}

\newcounter{jlisti}

\begin{document}

\title{Long time asymptotics for homoenergetic solutions of the Boltzmann equation.  Collision-dominated case.}

\author{Richard D. James \thanks{\emailrichard} , Alessia Nota \thanks{\emailalessia} , Juan J. L. Vel\'azquez \thanks{\emailjuan} \\[1em]
$\,^*$\UMaddress \\[0.5em]
$ ^\dag,^{\ddag}$\UBaddress}

\date{\today}

\newcommand{\email}[1]{E-mail: \tt #1}
\newcommand{\emailrichard}{\email{james@umn.edu}}
\newcommand{\emailalessia}{\email{nota@iam.uni-bonn.de}}
\newcommand{\emailjuan}{\email{velazquez@iam.uni-bonn.de}}

\newcommand{\UBaddress}{\em University of Bonn, Institute for Applied Mathematics\\
\em Endenicher Allee 60, D-53115 Bonn, Germany}
\newcommand{\UMaddress}{\em Department of Aerospace Engineering and Mechanics, University of Minnesota \\
\em 107, Akerman Hall, Minneapolis, MN 55455, USA}

\maketitle

\begin{abstract}

In this paper we present a formal analysis of the long-time asymptotics of a particular class of solutions of
the Boltzmann equation, known as
homoenergetic solutions,    
 which have the form $f\left(  x,v,t\right)  =g\left(
v-L\left(  t\right)  x,t\right)  $ where $L\left(  t\right)  =A\left(
I+tA\right)  ^{-1}$ with the matrix $A$ describing a shear flow or a
dilatation or a combination of both. We began this study in \cite{JNV1}.  Homoenergetic solutions satisfy an integro-differential equation which contains, in addition to the classical Boltzmann collision operator, a linear hyperbolic term. 
 In \cite{JNV1} it has been proved rigorously
the existence of self-similar solutions which describe the change of the
average energy of the particles of the system in the case in which there is a balance between the hyperbolic and the  collision term.

In this paper we focus in homoenergetic solutions for which the collision term is much larger than the hyperbolic term (collision-dominated behavior). 
In this case the long time asymptotics for the distribution of velocities is given by a time dependent Maxwellian distribution with changing temperature.  
\end{abstract}
\tableofcontents

\section{Introduction \label{sect1}}

In this paper we study the long time asymptotics of a class of non self-similar solutions of
the Boltzmann equation. 
The self-similar case was considered in  \cite{JNV1}. The class of solutions under consideration is motivated by
an invariant manifold of solutions of the equations of classical molecular
dynamics with certain symmetry properties (\cite{md, viscometry}).

As background, we recall briefly the properties of this manifold which have been summarized
also in \cite{JNV1}. Given a matrix $A\in M_{3\times3}\left(  \mathbb{R}%
\right)  $, and linearly independent vectors $e_{1},e_{2},e_{3}$ in
$\mathbb{R}^{3}$, we consider a time interval $[0,a)$ such that $\det(I+tA)>0$
for $t\in\lbrack0,a)$ with $a>0$. We consider any number of atoms labeled
$1,\dots,M$ with positive masses $m_{1},\dots,m_{M}$ and initial conditions
\begin{equation}
{y}_{k}(0)={y}_{k}^{0},\quad\dot{y}_{k}(0)={v}_{k}^{0},\quad k=1,\dots,M. \label{md_ic}
\end{equation}

These $M$ atoms will be called \textit{simulated atoms}. The simulated atoms
will be subject to the equations of molecular dynamics (to be stated
presently) with the initial conditions (\ref{md_ic}), yielding solutions
$y_{k}(t)\in\mathbb{R}^{3},\ 0\leq t<a,\ k=1,\dots,M$. In addition there will
be \textit{non-simulated atoms} with time-dependent positions $y_{\nu,k}(t)$,
indexed by a triple of integers $\nu=(\nu_{1},\nu_{2},\nu_{3})\in
\mathbb{Z}^{3},\ \nu\neq(0,0,0)$ and $k=1,\dots,M$. The nonsimulated atom
$(\nu,k)$ will have mass $m_{k}$. The positions of the nonsimulated atoms will
be given by the following explicit formulas based on the positions of the
simulated atoms:
\begin{equation}
y_{\nu,k}(t)=y_{k}(t)+(I+tA)(\nu_{1}e_{1}+\nu_{2}e_{2}+\nu_{3}e_{3}),\quad
\nu=(\nu_{1},\nu_{2},\nu_{3})\in\mathbb{Z}^{3},\ k=1,\dots,M. \label{nonsim}%
\end{equation}
For $k=1,\dots,M$ we denote as $f_{k}:\cdots\mathbb{R}^{3}\times\mathbb{R}%
^{3}\times{\mathbb{R}^{3}}\cdots\rightarrow\mathbb{R}$ the force on simulated
atom $k$. The force on simulated atom $k$ depends on the positions of all the
atoms. We assume that the force $f_{k}$ satisfies the usual conditions of
frame-indifference and permutation invariance \cite{md}. Then, the dynamics of
the simulated atoms is given by:
\begin{align}
m_{k}\ddot{y}_{k}  &  =f_{k}(\dots,y_{\nu_{1},1},\dots,y_{\nu_{1},M}%
,\dots,y_{\nu_{2},1},\dots,y_{\nu_{2},M},\dots),\label{md_sim}\\
{y}_{k}(0)  &  ={y}_{k}^{0},\quad\dot{y}_{k}(0)={v}_{k}^{0},\quad
k=1,\dots,M.\nonumber
\end{align}

\begin{figure}[th]
\label{fig1}\centering
\includegraphics [scale=0.3]{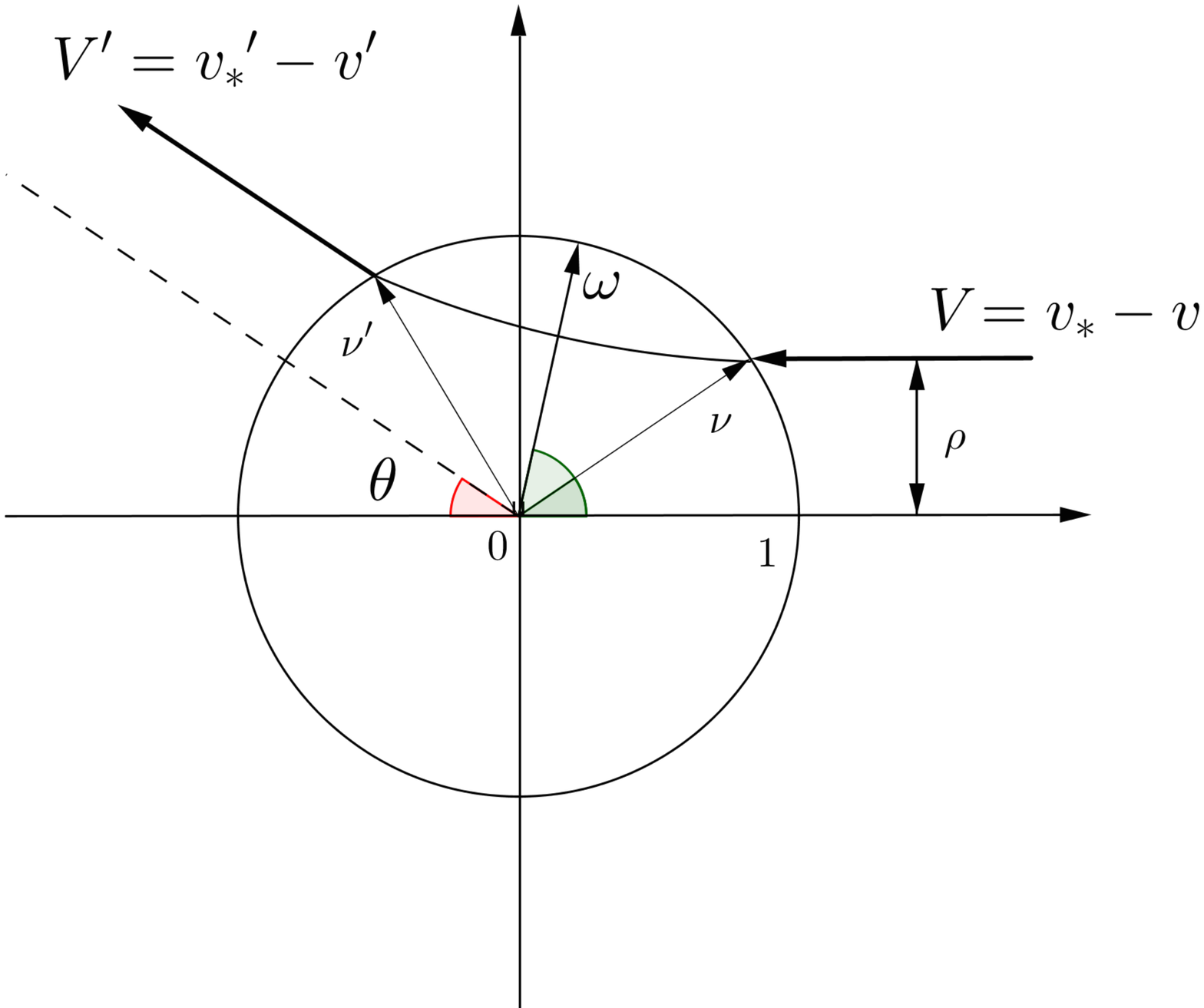}\caption{The two-body scattering.
The solution of the two-body problem lies in a plane, which is taken to be the
plane of the page, and the motion of molecule $\ast$ is plotted relative to
the unstarred molecule. The scalar $\rho$ is the impact parameter expressed in
microscopic units, $\rho\in[-1,1]$, and $\theta=\theta(\rho, |V|)$ is the
scattering angle. The scattering vector of (\ref{CM1}), (\ref{CM2}) is the
unit vector $\omega=\omega(v,V)$.} %
\end{figure}

Notice that these equations can be reduced to standard ODEs for the motions of
the simulated atoms substituting the formulas (\ref{nonsim}) into the right
hand side of (\ref{md_sim}). It is shown in \cite{md} and \cite{viscometry}
that in spite of the fact that the motions of the nonsimulated atoms are only
given by the formulas (\ref{nonsim}), the equations of molecular dynamics
(\ref{md_sim}) are exactly satisfied for each nonsimulated atom. Further information is also given in \cite{JNV1}.
 As discussed there, these results on molecular dynamics have a simple
interpretation in terms of the molecular density function of the kinetic
theory. To explain this interpretation we recall that the classical Boltzmann equation has the form
\begin{align}
\partial_{t}f+v\partial_{x}f  &  =\mathbb{C}f\left(  v\right)
\ \ ,\ \ f=f\left(  t,x,v\right) \nonumber\\
\mathbb{C}f\left(  v\right)   &  =\int_{\mathbb{R}^{3}}dv_{\ast}\int_{S^{2}%
}d\omega B\left(  n\cdot\omega,\left\vert v-v_{\ast}\right\vert \right)
\left[  f^{\prime}f_{\ast}^{\prime}-f_{\ast}f\right] , \ \label{A0_0}%
\end{align}
where $S^{2}$ is the unit sphere in $\mathbb{R}^{3}$ and $n=n\left(
v,v_{\ast}\right)  =\frac{\left(  v-v_{\ast}\right)  }{\left\vert v-v_{\ast
}\right\vert }.$ Here $(v,v_{\ast})$ is a pair of velocities in incoming
collision configuration (see Figure \ref{fig1}) and $(v^{\prime},v_{\ast
}^{\prime})$ is the corresponding pair of outgoing velocities defined by the
collision rule
\begin{align}
v^{\prime}  &  =v+\left(  \left(  v_{\ast}-v\right)  \cdot\omega\right)
\omega,\label{CM1}\\
v_{\ast}^{\prime}  &  =v_{\ast}-\left(  \left(  v_{\ast}-v\right)  \cdot
\omega\right)  \omega. \label{CM2}%
\end{align}
The unit vector $\omega=\omega(v,V)$ bisects the angle between the incoming
relative velocity $V=v_{\ast}-v$ and the outgoing relative velocity
$V^{\prime}=v_{\ast}^{\prime}-v^{\prime}$ as specified in Figure \ref{fig1}.
The collision kernel $B\left(  n\cdot\omega,\left\vert v-v_{\ast}\right\vert
\right)  $ is proportional to the cross section for the scattering problem
associated to the collision between two particles. We use the conventional
notation in kinetic theory, $f=f\left( t, x, v\right)  ,\ f_{\ast}=f\left(  t,
x, v_{\ast}\right)  ,\ f^{\prime}=f\left(  t, x, v^{\prime}\right)
,\ \ f_{\ast}^{\prime}=f\left(  t, x, v_{\ast}^{\prime}\right)  $.

We will assume that the kernel $B$ is homogeneous with respect to the variable
$\left\vert v-v_{\ast}\right\vert $ and we will denote its homogeneity by
$\gamma,\ $i.e.,
\begin{equation}
B\left(  n  \cdot\omega,\lambda\left\vert v-v_{\ast}\right\vert \right)  =\lambda
^{\gamma}B\left( n \cdot\omega,\left\vert v-v_{\ast}\right\vert \right)
,\ \ \lambda>0. \label{S8E7}%
\end{equation}

It is possible to find solutions of the Boltzmann equation \eqref{A0_0}
in the same spirit of the molecular dynamics simulation for discrete systems described above (see \eqref{nonsim}). 
Indeed, let us consider a ball $B_{r}(x)$ of any radius $r$ centered at
$x=(I+tA)(\nu_{1}e_{1}+\nu_{2}e_{2}+\nu_{3}e_{3})$, $(\nu_{1},\nu_{2},\nu
_{3})\in\mathbb{Z}^{3}$. The ansatz (\ref{nonsim}) implies that, the
velocities of the atoms in the ball $B_{r}(x)$ are determined by those in the
ball $B_{r}(0)$ by the time derivative of \eqref{nonsim}. The molecular density function $f(t,x,v)$ of the kinetic
theory describes the probability density of finding velocities in the small
neighborhood of a point $x$ at time $t$. Thus, the ansatz above for the
particle velocities in the balls $B_{r}(x)$ can be written down using
(\ref{nonsim}) and its time-derivative as:
\begin{equation}
f(t,x,v)=g(t,v-A(I+tA)^{-1}x).\label{ansatz}%
\end{equation}
The term $A(I+tA)^{-1}$ arises from conversion to the Eulerian form of the
kinetic theory.

The study of solutions of kinetic equations with the form (\ref{ansatz})
is\ also interesting from the general perspective of non-equilibrium
statistical mechanics. We have shown in \cite{JNV1} that for broad classes of
choices of $A$, there exist solutions of the Boltzmann equation satisfying
(\ref{ansatz}). We have obtained also explicit formulas for the entropy of some of these
solutions in terms of the time-dependent temperature and density.

An alternative viewpoint based on the theory of \textit{equidispersive
solutions} and leading to the same result is presented in Section \ref{hom}.
These are solutions of the Boltzmann equation with the form%
\begin{equation}
f\left(  t,x,v\right)  =g\left(  t,w\right)  \text{ \ \ with }w=v-\xi\left(
t,x\right)  .\ \label{HomSol}%
\end{equation}

Under mild smoothness conditions, solutions with the form
(\ref{HomSol}) exist if and only if $\xi(t,x)=A(I+tA)^{-1}x$. Formally, if $f$
is a solution of the Boltzmann equation (\ref{A0_0}) of the form
(\ref{ansatz}) the function $g$ satisfies
\begin{equation}
\partial_{t}g-\big(L\left(  t\right)  w\big)\cdot\partial_{w}g=\mathbb{C}g\left(
w\right)  \label{D1_0}%
\end{equation}
where the collision operator $\mathbb{C}$ is defined as in \eqref{A0_0}. These
solutions are called \textit{homoenergetic solutions} and were introduced by
Galkin \cite{Galkin1} and Truesdell \cite{T}.

Homoenergetic solutions of the Boltzmann equation have been studied in
\cite{Bobylev75}, \cite{Bobylev76}, \cite{BobCarSp}, \cite{CercArchive},
\cite{Cerc2000}, \cite{Cerc2002}, \cite{Galkin1}, \cite{Galkin2}, \cite{Galkin3}, \cite{garzo},
\cite{Nikol1}, \cite{Nikol2}, \cite{T}, \cite{TM}. Details about the precise contents of these papers will be given later in the corresponding sections where related results appear.

The properties of the solutions of (\ref{D1_0}) for large time $t$ depends
sensitively on the homogeneity of the kernel yielding the cross section of the
collision operator $\mathbb{C}g.$ In \cite{JNV1} we have focused on
the analysis of solutions of (\ref{D1_0}) for which the terms $L\left(
t\right)  w\cdot\partial_{w}g$ and $\mathbb{C}g\left(  w\right)$ are comparable.  This
happens for most choices of the matrix $A$ if the homogeneity $\gamma=\frac{\nu-5}{\nu-1}$ of the cross
section appearing in the operator $\mathbb{C}g\left(  w\right)  $ is zero, that is,
for potentials of the form $V\left(  x\right)
=\frac{1}{\left\vert x\right\vert ^{\nu-1}}$ with $\nu = 5$.  It
is customary in this case to say that the particles described by the
distribution $f$ in (\ref{HomSol}) are {\it Maxwell molecules}.

The main result that has been obtained in \cite{JNV1} is the rigorous proof
of existence of self-similar solutions in the class of homoenergetic flows if
the collision kernel describes the interaction between Maxwell molecules. In
all the cases when such self-similar solutions exist, the terms $L\left(
t\right)  w\cdot\partial_{w}g$ and $\mathbb{C}g\left(  w\right)  $ have a
comparable size as $t\rightarrow\infty.$

In this paper we will focus in the analysis of the possible long time
asymptotics of the solutions of (\ref{D1_0}) in the cases in which the
collision kernel describes the interactions between non Maxwellian molecules.
This behavior strongly depends on the homogeneity of the collision kernel $B$
and on the particular form of the hyperbolic terms in the equation for the
equidispersive flows, namely $L\left(  t\right)  w\cdot\partial_{w}g$. We will
see that depending on the homogeneity of the kernel we will have different
possible solutions of the Boltzmann equations for large
time. We have solutions for which the collision term $\mathbb{C}g\left(
w\right)  $ becomes the largest one as $t\rightarrow\infty.$ These solutions
are approximately Maxwellians, with a time-dependent temperature. The
differential equations which describe the evolution of the temperature can be
obtained by means of a suitable adaptation of the standard Hilbert expansion. 
The description of this family of solutions, that we will denote as the {\it collision-dominated case}, is the main content of this paper.

Conversely, there are also choices of $L\left(  t\right)  $ and collision kernels $B$ for
which the scaling properties of the different terms imply that the hyperbolic
terms are much more important than the collision terms. We will refer to these solutions as hyperbolic-dominated case. This case is discussed in \cite{JNV3}. 

The molecular dynamic simulation method described above can be rephrased as an
invariant manifold of the equations of molecular dynamics.  Our existence result in \cite{JNV1}, and other
results in  \cite{JNV3} and this paper
show that this manifold is inherited faithfully by the Boltzmann equation.  Our long term hope is
to be able to write a relatively simple but general asymptotic statistics on this manifold.   
One could conjecture that the situation is like the equilibrium
case, where the relevant manifold is $\calH = const.$ ($\calH$ is the Hamiltonian), the 
``statistics'' is the Maxwellian distribution (or, more generally, the Gibbs measure), 
and macroscopic properties are obtained as moments.  Taken together,  
our results of \cite{JNV1}, \cite{JNV3} and this paper on the dichotomy between hyperbolic-dominated
and collision-dominated behavior suggest that our asymptotic statistics is quite simple,
but not governed by single distribution as in the equilibrium case.   In particular, the selection
of asymptotic distribution is sensitively dependent on the growth (i.e., repulsiveness) of the
atomic forces.

The plan of the paper is the following. In Section \ref{hom} 
we recall the main properties of homoenergetic solutions of the Boltzmann
equation which have been obtained in \cite{JNV1}. In Section \ref{sec:wellpos} we describe general conditions under which the homoenergetic
solutions exist. Section \ref{sec:Hilbert} describes several
choices of matrices $L\left(  t\right)  $ and collision kernels $B$ for which
the long time asymptotics of the solutions is given by means of Hilbert
expansions, or more precisely perturbations of the Maxwellian distribution
with changing temperature. This section contains first a general theorem
yielding Hilbert expansions for a large class of matrices $L\left(  t\right)
$ and collision kernels $B.$ This abstract result is then applied to specific
choices of homoenergetic flows.  
In Section \ref{sec:tableconcl}, we summarize the
main results obtained in this paper, together with those  given in \cite{JNV1}, \cite{JNV3}, and we write
some concluding remarks.


\section{Homoenergetic solutions of the Boltzmann equation}
\label{hom}

We consider the molecular density function $f\left(  t,x,v\right)  $ solution
of \eqref{A0_0}, i.e. of
\begin{align*}
\partial_{t}f+v\partial_{x}f  &  =\mathbb{C}f\left(  v\right)
\ \ ,\ \ f=f\left(  t,x,v\right) \\\mathbb{C
}f\left(  v\right)   &  =\int_{\mathbb{R}^{3}}dv_{\ast}\int_{S^{2}}d\omega
B\left(  n\cdot\omega,\left\vert v-v_{\ast}\right\vert \right)  \left[
f^{\prime}f_{\ast}^{\prime}-f_{\ast}f\right] . \
\end{align*}
Formally, we can compute the density $\rho$, the average velocity $V$ and the internal
energy $\varepsilon$ at each point $x$ and time $t$ by means of%
\begin{equation}
\rho\left(  t,x\right)  =\int_{\mathbb{R}^{3}}f\left(  t,x,v\right)
dv,\ \ \rho\left(  t,x\right)  V\left(  t,x\right)  =\int_{\mathbb{R}^{3}%
}f\left(  t,x,v\right)  v\, dv. \label{S8E3}%
\end{equation}
The internal energy $\varepsilon\left(  t,x\right)  $ (or temperature) is
given by%
\[
\rho\left(  t,x\right)  \varepsilon\left(  t,x\right)  =\int_{\mathbb{R}^{3}%
}f\left(  t,x,v\right)  \left(  v-V\left(  t,x\right)  \right)  ^{2}dv.
\]

Homoenergetic solutions of (\ref{A0_0})  defined in \cite{Galkin1} and
\cite{T} (cf., also \cite{TM}) are solutions of the Boltzmann equation having
the form%
\begin{equation}
f\left(  t,x,v\right)  =g\left(  t,w\right)  \text{ \ \ with }w=v-\xi\left(
t,x\right) . \label{B1_0}%
\end{equation}

Notice that, under suitable integrability conditions, every solution of
(\ref{A0_0}) with the form (\ref{B1_0}) yields only time-dependent internal
energy and density%
\begin{equation}
\varepsilon\left(  t,x\right)  =\varepsilon\left(  t\right)  \ \ ,\ \ \rho
\left(  t,x\right)  =\rho\left(  t\right) . \label{eq:Hom1}%
\end{equation}
However, we have $V\left(  t,x\right)  =\xi\left(  t,x\right)  $ and therefore
the average velocity depends also on the position.

A direct computation shows that in order to have solutions of (\ref{A0_0})
with the form (\ref{B1_0}) for a sufficiently large class of initial data we
must have%
\begin{equation}
\frac{\partial\xi_{k}}{\partial x_{j}}\text{ independent on }x\text{ and
}\partial_{t}\xi+\xi\cdot\nabla\xi=0 . \label{B2_0}%
\end{equation}

The first condition implies that $\xi$ is an affine function of $x$. However,
we will restrict attention in this paper to the case in which $\xi$ is a
linear function of $x,$ for simplicity, whence%
\begin{equation}
\xi\left(  t,x\right)  =L\left(  t\right)  x, \ \label{B4_0}%
\end{equation}
where $L\left(  t\right)  \in M_{3\times3}\left(  \mathbb{R}\right)  $ is a
$3\times3$ real matrix. The second condition in (\ref{B2_0}) then implies that%
\begin{equation}
\frac{dL\left(  t\right)  }{dt}+\left(  L\left(  t\right)  \right)  ^{2}=0,
\quad L (0) = A, \label{B3_0}%
\end{equation}
where we have added an initial condition.

Classical ODE theory shows that there is a unique continuous solution of
(\ref{B3_0}),
\begin{equation}
L\left(  t\right)  =\left(  I+tA\right)  ^{-1}A=A\left(  I+tA\right)
^{-1},\ \label{B7_0}%
\end{equation}
defined on a maximal interval of existence $[0,a)$. On the interval $[0,a)$,
$\det\left(  I+tA \right) >0 $.

\subsection{Classification of homoenergetic solutions defined for arbitrary
large times. \label{ss:classeqsol}}

\bigskip

In this Section we recall the classification of homoenergetic flows which has
been obtained in \cite{JNV1} (cf., Theorem 3.1). More precisely, we describe
the long time asymptotics of $\xi\left(  t,x\right)  =L\left(  t\right)
x=\left(  I+tA\right)  ^{-1}Ax$ (cf. (\ref{B4_0}) and (\ref{B7_0})). As we
already discussed in \cite{JNV1} we observe that there are interesting
choices of $A\in M_{3\times3}\left(  \mathbb{R}\right)  $ for which $L\left(
t\right)  $ blows up in finite time, but we will restrict attention in this
paper to the case in which the matrix $\det(I + t A) >0$ for all $t \ge0$.

\begin{theorem}
[cf., Theorem 3.1 in \cite{JNV1}]\label{ClassHomEne} Let $A \in M_{3 \times
3}(\mathbb{R}) $ satisfy $\det(I + t A) >0$ for $t\ge0$ and let $L(t) = (I +
tA)^{-1}A$. Assume $L$ does not vanish identically. Then, there is an
orthonormal basis (possibly different in each case) such that the matrix of
$L(t)$ in this basis has one of the following forms:

\vspace{2mm} \noindent Case (i) Homogeneous dilatation:
\begin{equation}
L(t) = \frac{1}{t} I + O\bigg( \frac{1}{t^{2}} \bigg) \quad\mathrm{as}\ t
\to\infty.\label{T1E1}%
\end{equation}
\noindent Case (ii) Cylindrical dilatation (K=0), or Case (iii) Cylindrical
dilatation and shear ($K \ne0$):
\begin{equation}
L(t) = \frac{1}{t} \left(
\begin{array}
[c]{ccc}%
1 & 0 & K\\
0 & 1 & 0\\
0 & 0 & 0
\end{array}
\right)  + O\bigg( \frac{1}{t^{2}} \bigg) \quad\mathrm{as}\ t \to
\infty.\label{T1E2}%
\end{equation}
\noindent Case (iv). Planar shear:
\begin{equation}
L(t) = \frac{1}{t} \left(
\begin{array}
[c]{ccc}%
0 & 0 & 0\\
0 & 0 & K\\
0 & 0 & 1
\end{array}
\right)  + O\bigg( \frac{1}{t^{2}} \bigg) \quad\mathrm{as}\ t \to
\infty.\label{T1E3}%
\end{equation}
\noindent Case (v). Simple shear:
\begin{equation}
L(t) = \left(
\begin{array}
[c]{ccc}%
0 & K & 0\\
0 & 0 & 0\\
0 & 0 & 0
\end{array}
\right)  , \quad K \ne0.\label{T1E5}%
\end{equation}
\noindent Case (vi). Simple shear with decaying planar dilatation/shear:
\begin{equation}
L(t) = \left(
\begin{array}
[c]{ccc}%
0 & K_{2} & 0\\
0 & 0 & 0\\
0 & 0 & 0
\end{array}
\right)  + \frac{1}{t} \left(
\begin{array}
[c]{ccc}%
0 & K_{1} K_{3} & K_{1}\\
0 & 0 & 0\\
0 & K_{3} & 1
\end{array}
\right)  + O\bigg( \frac{1}{t^{2}} \bigg) , \quad K_{2} \ne0.\label{T1E6}%
\end{equation}
\noindent Case (vii). Combined orthogonal shear:
\begin{equation}
L(t) = \left(
\begin{array}
[c]{ccc}%
0 & K_{3} & K_{2} - t K_{1} K_{3}\\
0 & 0 & K_{1}\\
0 & 0 & 0
\end{array}
\right) , \quad K_{1} K_{3} \ne0.\label{T1E4}%
\end{equation}
\label{ClassHomEne}
\end{theorem}

\subsection{Behavior of the density and internal energy for homoenergetic
solutions}

\bigskip

Our goal is to to construct solutions of \eqref{D1_0} with the different choices of $L\left(  t\right)  $ obtained in Theorem \ref{ClassHomEne}.  
The equation describing homoenergetic flows (cf. \eqref{D1_0}) reads as
\begin{equation}
\partial_{t}g-\big(L\left(  t\right)  w\big)\cdot\partial_{w}g=\mathbb{C}g\left(
w\right)  \ \label{S8E9}%
\end{equation}
where the kernel $B$ in the collision operator is homogeneous
with homogeneity $\gamma$ (cf. (\ref{S8E7})). 

The solutions in which we are interested have certain scaling properties. Two quantities which play a crucial role determining these rescalings are the density $\rho\left(  t\right)  $ and the
internal energy $\varepsilon\left(  t\right) $ which in the case of homoenergetic solutions are given by (cf.
(\ref{S8E3})):%
\begin{equation}
\rho\left(  t\right)  =\int_{\mathbb{R}^{3}}g\left(  t,dw\right)
\ \ ,\ \ \ \varepsilon\left(  t\right)  =\int_{\mathbb{R}^{3}}\left\vert
w\right\vert ^{2}g\left(  t,dw\right).  \label{S8E4}%
\end{equation}
We note that these two quantities will be finite for each $t>0$ for all the solutions
considered in this paper.

We will need to describe the time evolution of the density and the internal energy. 
Integrating (\ref{D1_0}) with respect to the velocity variable and using the conservation of mass property of the
collision kernel, we obtain an evolution equation for the density,
\begin{equation}
\partial_{t}\rho\left(  t\right)  +\Tr \left(  L\left(  t\right)  \right)
\rho\left(  t\right)  =0, \label{S8E5}%
\end{equation}
whence%
\begin{equation}
\rho\left(  t\right)  =\rho\left(  0\right)  \exp\left(  -\int_{0}%
^{t}\Tr \left(  L\left(  s\right)  \right)  ds\right)  . \label{S8E6}%
\end{equation}

For the
internal energy $\varepsilon\left(  t\right),$  it is not possible to derive a similarly simple evolution equation because the term $-\big(L\left(
t\right)w\big) \cdot\partial_{w}g$ on the left-hand side of (\ref{D1_0}) yields, in
general, terms which cannot be written only in terms of $\rho\left(
t\right)  ,\ \varepsilon \left(  t \right).$  This is the closure problem of the general system 
of equations of moments of the kinetic theory. 
 Actually, these terms have an
interesting physical meaning, because they produce heating or cooling of the
system and therefore they contribute to the change of $\varepsilon \left( t \right).$ 
To obtain the precise form of these terms we need to study the
detailed form of the solutions of (\ref{D1_0}). The rate of growth or decay of
$\varepsilon\left(  t\right)  $ would then typically appear as an eigenvalue
of the corresponding PDE problem.
\bigskip

Observe that the particle density $\rho(t)$ in \eqref{S8E6} is not necessarily constant. It will be convenient in the following to reformulate \eqref{S8E9} in a form in which the particle density is constant. To this end we introduce a new function $\tilde{g}$ by means of
$$g(t, w)=\tilde{g}(t,w) \exp\left(  -\int_{0}%
^{t}\Tr \left(  L\left(  s\right)  \right)  ds\right).$$ 
Then, using \eqref{S8E9}, $\tilde{g}$ satisfies
\begin{equation}\label{eq:homgtilde}
\partial_{t}\tilde{g}-\partial_{w}\cdot \big(L\left(  t\right)  w \tilde{g} \big)=\exp\left(  -\int_{0}%
^{t}\Tr \left(  L\left(  s\right)  \right)  ds\right)\mathbb{C}\tilde{g}\left(
w\right) .
\end{equation}
Notice that in all the cases described in Theorem \ref{ClassHomEne} we have 
\begin{equation}\label{eq:Lt}
L\left(  t\right)\sim l(t)L_0+o(l(t))
\end{equation}
where $l(t)$ is either $(t+1),\,1$ or $\frac{1}{(t+1)}$ and $L_0\in M_{3\times 3}(\R)$, $L_0\neq 0$. We can reformulate \eqref{eq:homgtilde} using the new time variable $\tau $ defined as 
$\tau=\int_{0}^{t} l(s)ds$. Note that $t\to \tau$ defines a strictly monotone mapping from $[0,\infty)$ to $[0,\infty)$ and in particular $\tau \to \infty$ as $t\to \infty$.  Then \eqref{eq:homgtilde} becomes
 \begin{equation}\label{eq:homgtilde_2}
\partial_{\tau}\tilde{g}-\partial_{w}\cdot \big(Q\left(  \tau\right)  w \tilde{g} \big)=\mu(\tau)\mathbb{C}\tilde{g}\left(w\right) 
\end{equation}
where 
\begin{equation}\label{eq:defmu}
\mu(\tau)=\frac{1}{l(t)} \exp\left(  -\int_{0}^{t}\Tr \left(  L\left(  s\right)  \right)  ds\right)
\end{equation} 
and 
\begin{equation}\label{eq:defQ}
Q(\tau)=L_0+o(1),\quad\text{as} \;\; \tau \to \infty.
\end{equation}  

\section{Well posedness theory for homoenergetic flows} \label{sec:wellpos}

We recall here some results on the well posedness for homoenergetic flows with the form (\ref{B1_0}), (\ref{B4_0}), (\ref{B3_0}). 
This issue has been addressed by Cercignani in \cite{CercArchive},
\cite{Cerc2000} who proved, using the $L^{1}$ theory for the Boltzmann
equation, that homoenergetic flows, in the case of simple shear, exist for a
large class of initial data $g_{0}\left(  w\right)$. On the other hand, in
\cite{JNV1} we proved a well posedness results in the class of Radon
measures for more general choices of the function $L(t)$ 
in the case of collision kernel associated to Maxwell molecules (i.e.,
homogeneity $\gamma=0$).

We follow here the strategy proposed in \cite{JNV1} and we study the well-posedness of \eqref{eq:homgtilde_2} which we rewrite here replacing $\tilde{g}$ by $g$ and $\tau$ by $t$ for the sake of simplicity. We then have:
\begin{align}
\partial_{t}g-\partial_{w}\cdot\left(  \left[  Q(t)w\right]  g\right)   &
=\mu(t) \mathbb{C}g\left(  w\right) \label{eq:boltzgen}\\
\mathbb{C}g\left(  w\right)   &  =\int_{\mathbb{R}^{3}}dw_{\ast}\int_{S^{2}%
}d\omega B\left(  n\cdot\omega,\left\vert w-w_{\ast}\right\vert \right)
\left[  g^{\prime}g_{\ast}^{\prime}-g_{\ast}g\right]
\ ,\label{eq:collboltzgen}\\
g\left(  0,w\right)   &  =g_{0}\left(  w\right) , \ \label{eq:invalBol}%
\end{align}
where we impose the following growth conditions which cover all the cases considered in this paper:
\begin{equation}\label{eq:cdtmu}
c_1 e^{-A t}\leq \mu(\tau)\leq c_2 e^{A t}, \quad \text{ with }\;\; 0<c_1,\,c_2, \,A <\infty ,
\end{equation} 
and 
\begin{equation}
Q(\cdot)\in C^{1}\left(  \left[  0,\infty\right)  ;M_{3\times3}\left(
\mathbb{R}\right)  \right)  \ \ ,\ \ \left\Vert Q(t)\right\Vert \leq
c_{3}e^{At},\text{ with } 0<c_3, \,A <\infty, \label{T3E2}%
\end{equation}
with the norm $\left\Vert \cdot\right\Vert $ in $M_{3\times3}\left(
\mathbb{R}\right)  $ :%
\begin{equation}
\left\Vert M\right\Vert =\max_{i,j}\left\Vert m_{i,j}\right\Vert
\ \ \text{with }M=\left(  m_{i,j}\right)  _{i,j=1,2,3} . \label{T3E1}%
\end{equation}


For the sake of completeness we now recall some definitions and notation used
in \cite{JNV1} in order to prove well-posedness results. We set $\mathcal{M}_{+}\left(  \mathbb{R}_{c}^{3}\right)  $ to be the set of
Radon measures in $\mathbb{R}_{c}^{3}$ which denotes the compactification of
$\mathbb{R}^{3}$ by means of a single point $\infty$. This is a technical
issue that we need in order to have convenient compactness properties for some
subsets of $\mathcal{M}_{+}\left(  \mathbb{R}_{c}^{3}\right)  .$ The space
$C\left(  \left[  0,\infty\right)  :\mathcal{M}_{+}\left(  \mathbb{R}_{c}%
^{3}\right)  \right)  $ is defined endowing $\mathcal{M}_{+}\left(
\mathbb{R}_{c}^{3}\right)  $ with the measure norm
\begin{equation}
\left\Vert \nu\right\Vert _{M}=\sup_{\varphi\in C\left(  \mathbb{R}_{c}%
^{3}\right)  :\left\Vert \varphi\right\Vert _{\infty}=1}\left\vert \nu\left(
\varphi\right)  \right\vert =\int_{\mathbb{R}_{c}^{3}} \varphi \, d \nu (  w)  . \label{T3E2b}%
\end{equation}
Note that this definition implies that the total measure of $\mathbb{R}%
_{c}^{3}$ is finite if $\nu\in\mathcal{M}_{+}\left(  \mathbb{R}_{c}%
^{3}\right)  .$ 
Moreover $\varphi\in C\left(  \mathbb{R}_{c}^{3}\right)  $
implies that the limit value $\varphi\left(  \infty\right)  $ exists.

We use the following concept of weak solutions of 
(\ref{eq:boltzgen})-(\ref{eq:invalBol}).

\begin{definition}
\label{WeakSol}We will say that $g\in C\left(  \left[  0,\infty\right)
:\mathcal{M}_{+}\left(  \mathbb{R}_{c}^{3}\right)  \right)  $ is a weak
solution of (\ref{eq:boltzgen})-(\ref{eq:invalBol}) with initial value
$g\left(  0,\cdot\right)  =g_{0}\in\mathcal{M}_{+}\left(  \mathbb{R}_{c}%
^{3}\right)  $ if for any $T\in\left(  0,\infty\right)  $ and any test
function $\varphi\in C\left(  \left[  0,T\right)  :C^{1}\left(  \mathbb{R}%
_{c}^{3}\right)  \right)  $ the following identity holds
 \begin{align}\label{T3E6}
&  \int_{\mathbb{R}^{3}}\varphi\left(  T,w\right)  g\left(  T,dw\right)
-\int_{\mathbb{R}^{3}}\varphi\left(  0,w\right)  g_{0}\left(  dw\right)
\\&  
=-\int_{0}^{T}dt\int_{\mathbb{R}^{3}}\int\partial_{t}\varphi g\left(
t,dw\right)  -\int_{0}^{T}dt\int_{\mathbb{R}^{3}}\left[  Q(t)w\cdot
\partial_{w}\varphi\right]  g\left(  t,dw\right)  \nonumber\\
& \quad +\frac{\mu(t)}{2}\int_{0}^{T}dt\int_{\mathbb{R}^{3}}\int_{\mathbb{R}^{3}}%
\int_{S^{2}}d\omega g\left(  t,dw\right)  g\left(  t,dw_{\ast}\right)
B\left(  n\cdot\omega,\left\vert w-w_{\ast}\right\vert \right) 
\Big[\varphi\left(  t,w^{\prime}\right)  +\varphi\left(  t,w_{\ast}^{\prime
}\right)\nonumber \vspace{5mm}\\&  \qquad -\varphi\left(  t,w\right)  -\varphi\left(  t,w_{\ast}\right)
\Big]. \nonumber
\end{align}
\end{definition}

We will use the following norms:
\begin{equation}
\left\Vert g\right\Vert _{1,s}=\int_{\mathbb{R}^{3}}\left(  1+\left\vert
w\right\vert ^{s}\right)  g\left(  dw\right)  \ \ \text{for\ }g\in
\mathcal{M}_{+}\left(  \mathbb{R}_{c}^{3}\right)  \text{\ },\ \ s>0.
\label{NormS}%
\end{equation}

\bigskip 
In the following sections we will consider asymptotic expansions of solutions of \eqref{eq:boltzgen}-\eqref{eq:invalBol} as $t\to \infty$. In order to make rigorous these expansions the issue of the global well-posedness should be addressed.  Global existence of solutions of \eqref{eq:boltzgen}-\eqref{eq:invalBol} is expected for all reasonable collision kernels $B$ with arbitrary homogeneity $\gamma$. Although we will not make them rigorous in this paper, we indicate a typical  global well-posedness result that can be proved rigorously by adapting the current available methods for the homogeneous Boltzmann equation. For instance, the following result holds.

\begin{theorem}
\label{WellPos} 
Let $B\left(  n\cdot\omega,\left\vert v-v_{\ast}\right\vert \right)=\left\vert v-v_{\ast}\right\vert^{\gamma}b(\cos\theta)$ be a collision  kernel satisfying Grad's cut-off assumption 
$$\int_{0}^{\pi} b(\cos\theta) \sin\theta d\theta< \infty.$$

Suppose that $g_{0}\in\mathcal{M}_{+}\left(  \mathbb{R}%
^{3}\right)  $ satisfies, for $s = 2 $, 
\[
\left\Vert g_0\right\Vert _{1,s} <\infty. 
\]
Then, there exists a weak solution $g\in C\left(  \left[
0,\infty\right)  :\mathcal{M}_{+}\left(  \mathbb{R}_{c}^{3}\right)  \right)  $
in the sense of Definition \ref{WeakSol} to the initial value problem
(\ref{eq:boltzgen})-(\ref{eq:invalBol}) with $Q$ 
satisfying (\ref{T3E2}).
\end{theorem}

In particular, this theorem can be proved adapting the proof of Theorem 4.2 in \cite{JNV1},
using standard arguments from the theory of homogeneous Boltzmann equations as
described in \cite{CIP}, \cite{Desv}, \cite{He}, \cite{V02}.

Note that the contribution of the linear term $\partial_{w}\cdot\left(  \left[  Q(t)w\right]  G\right)$ in \eqref{eq:boltzgen} in the moment estimates is trivial. Therefore, the moment estimates for the solutions of \eqref{eq:boltzgen}-\eqref{eq:invalBol} are basically identical, for finite time, to the ones obtained for the  homogeneous Boltzmann equation.

\bigskip

\section{Hilbert expansions for homoenergetic flows with collision-dominated
behavior}\label{sec:Hilbert}


\subsection{Hilbert expansions for general homoenergetic flows}

\label{sec:genHilbexp} \bigskip

Our goal is to compute the asymptotic expansions for some solutions of  \eqref{eq:boltzgen}-\eqref{eq:invalBol}, as $t\to \infty$, in which the dominant term is the collision term.

The problem under consideration is the following generalized reduced Boltzmann equation \eqref{eq:boltzgen} which reads as
 \begin{equation*}
\partial_{t}g-\partial_{w}\cdot\left(  \left[  Q(t)w\right]  g\right) =\mu(t)\int
_{\mathbb{R}^{3}}dw_{\ast} \int_{S^{2}}d\omega\, B\left( n\cdot \omega,\left\vert
w-w_{\ast}\right\vert \right)  \left[  g^{\prime}g_{\ast}^{\prime}-g_{\ast
}g\right], 
\end{equation*}
where $B$ satisfies \eqref{S8E7} and $Q(t)$ is as in \eqref{eq:defQ}.

We remark  that the associated particle density $\rho(t)$ is constant if $g$ solves \eqref{eq:boltzgen}. Moreover, \eqref{eq:defQ} holds and $\mu(t)$ defined as in \eqref{eq:defmu} will converge to a constant or increase exponentially depending on the choice of $L(t)$ in Theorem \ref{ClassHomEne}.

In this section we choose values of the homogeneity parameter
$\gamma$ and the matrix $L(t)$ characterizing the homoenergetic flow so that the collision terms dominate the hyperbolic term.
In such a case the asymptotics of the velocity dispersion can be computed formally
by means of a suitable adaptation of the classical Hilbert expansions around the Maxwellian equilibrium. Since the collision terms are the dominant ones, we
expect that, in the long time asymptotics, the solutions should behave as a
Maxwellian distribution with increasing or decreasing temperature depending on
the sign of the homogeneity parameter $\gamma$.

In order to simplify the notation we introduce the bilinear form
\begin{equation}
\label{eq:symbilcoll}\mathbb{C}\left[  f,g\right]  \left(  w\right)  =\frac
{1}{2}\int_{\mathbb{R}^{3}}dw_{\ast}\int_{S^{2}}d\omega B\left(
n\cdot \omega,\left\vert w-w_{\ast}\right\vert \right)  \left[  f^{\prime}g_{\ast
}^{\prime}+g^{\prime}f_{\ast}^{\prime}-f_{\ast}g-g_{\ast}f\right]  .
\end{equation}
Then $\mathbb{C}g\left(  w\right)  =\mathbb{C}\left[  g,g\right]  \left(
w\right)  . $ 
Moreover, we also introduce the following operator
\begin{align}\label{eq:linL}
 \mathbb{L}\left[
H\right]  \left(  \xi\right) =\int_{\mathbb{R}^{3}}d\xi_{\ast}%
\int_{S^{2}}d\omega B\left(  n\cdot \omega,\left\vert \xi-\xi_{\ast}\right\vert
\right)  e^{-\left\vert \xi_{\ast}\right\vert ^{2}}\left[  H_{\ast}^{\prime
}+H^{\prime}-H-H_{\ast}\right],
\end{align}
for any $H\in \mathcal{D}(H)\subset L^{2}\left(  \mathbb{R}%
^{3};e^{-\left\vert \xi\right\vert ^{2}} d\xi\right)$. Note that the space $L^{2}\left(  \mathbb{R}%
^{3};e^{-\left\vert \xi\right\vert ^{2}} d\xi\right)  $ is a Hilbert space with the scalar
product
\begin{equation}
\label{eq:weightedscpd}\left\langle f,g\right\rangle _{w} =\int_{{\mathbb{R}%
}^{3}} f(\xi) g(\xi) e^{-|\xi|^{2}}d\xi.
\end{equation}

The operator $-\mathbb{L}$ 
is a well studied linear operator in kinetic theory. 
It is a nonnegative, self-adjoint operator which has good functional analysis properties for suitable choices of the collision kernel $B$.  
Its kernel consists of collision invariants, i.e. it is the subspace spanned by the functions $\left\{
1,\xi,\left\vert \xi\right\vert ^{2}\right\}  .$ 
We define the subspace $\mathcal{W}=\left\{
1,\xi,\left\vert \xi\right\vert ^{2}\right\} ^{\perp} \subset L^{2}\left(  \mathbb{R}%
^{3};e^{-\left\vert \xi\right\vert ^{2}} d\xi\right).$ For further details see we refer to \cite{CIP}, \cite{TM}. 
We remark that the invertibility of the operator $\mathbb{L}$ in the subspace $\mathcal{W}$ has been rigorously proved for several collision
kernels. (See for instance the discussion in \cite{TM}).

The main result of this Section is the following Conjecture.

\bigskip

\begin{conjecture}
\label{th:genHilbexp} 
Suppose that the cross-section $B$ satifies condition
\eqref{S8E7}. Then, under suitable assumptions on the homogeneity $\gamma$ and on $\mu(t)$, there exists $g\left(  \cdot\right)  \in C\left(  \left[
0,\infty\right]  :\mathcal{M}_{+}\left(  \mathbb{R}_{c}^{3}\right)  \right)  $, weak solution of the Boltzmann
equation \eqref{eq:boltzgen} in the sense of Definition \ref{WeakSol} and a function $\beta(t)$ such that the solution $g$ behaves like
a Maxwellian distribution for long times, i.e.,
\begin{equation}
\label{eq:MaxHilb}\beta(t)^{-\frac{3}{2}}g\big(t,\frac{\xi}{\sqrt{\beta(t)}%
}\big) \to C_0\, e^{-\left\vert \xi\right\vert ^{2}}\quad\text{in}\quad
L^{2}\left(  \mathbb{R}^{3};e^{-\left\vert \xi\right\vert ^{2}}d\xi\right) \quad \text{as}\quad t\to\infty,
\end{equation} 
with $C_0=\pi^{-\frac{3}{2}}$. 
More precisely, we have the following cases.

\begin{itemize}
\item[1)] 
Assume that 
\begin{equation}\label{eq:matrixQ}
Q(t)=L_0+ O\left(\frac{1}{t^{1+\delta}}\right)\quad\text{ with}\quad \delta>0.
\end{equation}
and 
\begin{equation}
\label{eq:case1Hilb}
\Tr(L_0)\neq0.
\end{equation}
We define $a:=  \frac{2}{3}\Tr(L_0)$. Then,
if $\mu(t)e^{-\frac{\gamma}{2}at}\gg t^{1+\delta}$ as $t\to\infty$, with $\delta>0$, the asymptotic behavior is given
by a Maxwellian distribution as in \eqref{eq:MaxHilb} with
\begin{equation}
\label{eq:beta1Hilb}\beta(t)=C\,e^{at}(1+o(1)) \quad \text{as}\quad t\to \infty,%
\end{equation}
where $C>0$ is a numerical constant.

\item[2)]
Let $\gamma>0$ and assume that 
\begin{equation}
\label{eq:case2Hilb}
\Tr(Q(t))=0
\end{equation}
where $Q(t)=L_0+o(1)$, as $t\to \infty$ with $L_0\neq 0$.  Define $\lambda(t)=\int_{0}^{t} \frac
{ds}{\mu(s)}$ and suppose that $\lim_{t\to\infty}\lambda(t)=+\infty$ and $\lim_{t\to\infty}\frac{\lambda'(t)}{\lambda(t)}=0$.
Assume also that 
\begin{equation}
\label{eq:GreenKubo}
b=\left\langle \xi\cdot L_0 \xi, (-\mathbb{L})^{-1}[ (\xi\cdot L_0\xi)] )
\right\rangle _{w}>0,
\end{equation}
 where the operator $\mathbb{L}$ is as in \eqref{eq:linL}.
Then the asymptotic behavior is given by
a Maxwellian distribution  as in \eqref{eq:MaxHilb}  with increasing temperature,
and $\beta(t)$ satisfies
\begin{equation}
\label{eq:beta2Hilb}
\beta(t) = \left( \frac{4}{3} \gamma\, b \lambda(t)\right) ^{-\frac{2}{\gamma}}(1+o(1)) \quad\text{as}\quad t\to\infty.
\end{equation}
\end{itemize}
\end{conjecture}

\medskip 

\begin{remark}
In the case 1) we emphasize that for $\mu(t)=1$ we must choose the homogeneity
$\gamma$ satisfying $a\cdot\gamma<0$ in order to obtain a dynamics dominated
by collisions.
\end{remark}
\begin{remark}
A typical example of a function $\mu(t)$ satisfying the assumptions of the case 2) is $\mu(t)=(t+1)^{r}$, with $0\leq r<1$. Notice that since the operator $-\mathbb{L}$ is nonnegative we might expect to have $b>0$ for a large class of collision kernels.
\end{remark}
\begin{remark}
We observe that the condition \eqref{eq:case2Hilb} implies that $\Tr(L_0)=0$. Then, an elementary computation shows that $\xi\cdot L_0 \xi \in \mathcal{W}$. Thus the constant $b$ in the case 2) is well defined.
\end{remark}
\begin{remark}
As we will see in detail in Section \ref{sec:HilbExp}, the cases (i), (ii), (iii) and (iv) of Theorem  \ref{ClassHomEne} can be reduced to case 1) of Conjecture \ref{th:genHilbexp}. The cases (v) and (vii) of Theorem  \ref{ClassHomEne} can be reduced to case 2) of Conjecture \ref{th:genHilbexp}. The case (vi) of Theorem  \ref{ClassHomEne} cannot be included in a strict sense in none of the two cases of the Conjecture \ref{th:genHilbexp} but it is possible to obtain the asymptotics of the homoenergetic solutions adapting the arguments in the justification of the case 2) of the Conjecture \ref{th:genHilbexp}. The details are discussed in Subsection \ref{ssec:sheardil}.
\end{remark}
\begin{remark}
We stress that, in order to obtain a rigorous proof of the Theorem above,  we
should control the remainder of the Hilbert expansion (i.e. the boundedness in
the $L^{2}$ norm or perhaps in suitable weighted norms). Also questions related to the invertibility of the operator $\mathbb{L}$ should be addressed. For details about these points for some kernels  we refer to \cite{CIP, TM}.
In this paper we will not try to address these issues to avoid cumbersome
technicalities but we will describe the formal computations which suggest that the Conjecture \ref{th:genHilbexp} is true for suitable collision kernels.  It is likely that the current available tools to prove rigorously the Hilbert Expansions could be adapted in order to provide a rigorous proof of Conjecture \ref{th:genHilbexp} under suitable assumptions on the collision kernels.
Detailed information on the rigorous proof of the Hilbert expansions and on the invertibility properties of the operator $\mathbb{L}$ can be found in
\cite{GS1, MS, SR}.
\end{remark}

\medskip 

\begin{proofof}
[Justification of Conjecture \ref{th:genHilbexp}] 

We can assume without loss of generality the normalization $\int
g\left( t,dw\right)  =1$, modifying the function $\mu(t)$ if needed.  
Using that $\int w\, \mathbb{C}  g\left(  t,dw\right)  =0$ as well as the fact that $\int w\, \partial_{w}\cdot\left(  \left[  Q(t)w\right]  g( t,dw)\right) = -Q(t)\int w\, g( t,dw)$  it follows that if $\int w g_0(dw)=0$ than $\int w g(t, dw)=0$ for any $t>0$. We can then assume without loss of generality that 
$\int w\, g\left( t, dw\right)  =0.$ We look for a solution of \eqref{eq:boltzgen} with the form
\begin{equation}
g( t, w ) \sim C_{0}\left(  \beta\left(  t\right)  \right)  ^{\frac{3}{2}}\exp\left(
-\beta\left(  t\right)  \left\vert w\right\vert ^{2}\right)  \left[
1+h_{1}\left( t, w\right)  +h_{2}\left( t,w \right)  ...\right]
\ \label{H1E1}
\end{equation} 
with 
\begin{equation}\label{eq:orderhk}
 1 \gg \vert h_1\vert \gg \vert h_2\vert  \quad \text{as}\quad t\to \infty.
\end{equation} In order to determine the function $\beta(t)$ we impose the orthogonality conditions 
\begin{equation}\label{eq:orthcdt}
\int\exp\left(  -\beta\left(  t\right)  \left\vert w\right\vert ^{2}\right)
h_{k}\left(  t, w\right) \psi(w) dw=0\ \ ,\ \ k=1,2,3,...
\end{equation}
with $\psi(w) \in\{1,w, \vert w\vert^2\}$. 
 Therefore, using that $\int g\left( t,dw \right) =1$ we obtain
\begin{equation}\label{eq:defC0}
C_0 \beta^{\frac{3}{2}} \int \exp\left(  -\beta(t)  \left\vert w\right\vert ^{2}\right)  dw  =1 \quad \text{whence}\quad C_0=\frac{1}{\pi^{\frac{3}{2}}}.
\end{equation}
Moreover, \eqref{eq:orthcdt}  
implies 
\begin{equation}\label{eq:defbeta}
\int\left\vert w\right\vert ^{2}g\left( t,dw\right) = C_0 \left(
\beta\left(  t\right)  \right)^{\frac{3}{2}} \int \left\vert w\right\vert ^{2} \exp\left(  -\beta\left(
t\right)  \left\vert w\right\vert ^{2}\right)  dw =\frac{3}{2}\frac{1}{ \beta(t)},
\end{equation}
which gives the definition of $\beta\left(  t\right)$.

Notice that the expansion is not exactly the usual Hilbert expansion because
there is not a small parameter $\varepsilon$ in this setting. Actually the
small parameter is $\frac{1}{t}$ (as $t\rightarrow\infty$).

It is convenient to use a group of variables for which the Maxwellians have
a constant temperature. We define: 
\begin{equation}
\label{eq:ansHilbgen}\xi=\sqrt{\beta\left(  t\right)  }w\ \ ,\ \ g\left(
t, w\right)  =\left(  \beta\left(  t\right)  \right)  ^{\frac{3}{2}}G\left(
t,\xi\right)
\end{equation}
Then \eqref{eq:boltzgen} becomes:  
\begin{equation}
\label{eq:ansHilbgen_1}
\partial_{t}G+\frac{\beta_{t}}{2\beta}\xi\cdot\partial_{\xi}G+\frac{3}{2}
\frac{\beta_{t}}{\beta}G-\partial_{\xi}\cdot(Q(t)\xi G)=\mu(t)\beta^{-\frac{\gamma}
{2}}\mathbb{C}\left[  G,G\right]  \left(  \xi\right)  .
\end{equation}
Notice that the factor $\beta^{-\frac{\gamma}{2}}$ is due to the scaling
properties of the cross-section $B$ given in \eqref{S8E7}.

We define $H_{k}\left(t,\xi\right)=h_{k}\left(t,w\right)$ for $k=1,2,\dots$. Therefore, 
the expansion (\ref{H1E1}) becomes
\begin{equation}
\label{H1E1_bis}G\left(  \xi,t\right)  =C_{0}\exp\left(  -\left\vert
\xi\right\vert ^{2}\right)  \left[  1+H_{1}\left(  \xi,t\right)  +H_{2}\left(
\xi,t\right)  ...\right]
\end{equation}
where $C_0$ is given by \eqref{eq:defC0} 
and \eqref{eq:orthcdt}  yields the orthogonality conditions 
\begin{equation}\label{const}
\int \exp\left(  -\left\vert \xi\right\vert ^{2}\right)  H_{k}\left(
t, \xi\right) \psi(\xi) d^{3}\xi   =0\ \ ,\ \ k=1,2,3,... 
\end{equation}
with $\psi(\xi) \in \{1,\xi,\vert \xi\vert^2\}$. 
 
Notice that the definition of the functions $H_k$ as well as \eqref{eq:orderhk} implies   
\begin{equation}\label{eq:cdtHk}
1\gg \vert H_{1}\vert \gg \vert H_{2}\vert \gg....\text{ as }t\rightarrow\infty .
\end{equation}

We set $H_0=1$. From now on we will write
\begin{equation}
\label{equivGwH}
G_{k}=C_{0}\exp\left(  -\left\vert \xi\right\vert ^{2}\right)H_{k}\ \ , \quad k=0,1,2,\dots
\end{equation}
Then $G=G_{0}+G_{1}+G_{2}+\dots$  and plugging this series into \eqref{eq:ansHilbgen_1} we obtain
\begin{align*}
&  \partial_{t}\big(G_{0}+G_{1}+G_{2}\big)+\frac{\beta_{t}}{2\beta}\xi \cdot
\partial_{\xi} \big(G_{0}+G_{1}+G_{2}\big) +\frac{3}{2}\frac{\beta_{t}}{\beta}
\big(G_{0}+G_{1}+G_{2}\big) -\partial_{\xi} \cdot \left(  Q(t)\xi\big(G_{0}+G_{1}%
+G_{2}\big)\right) +\dots \\
&  =\mu(t)\left( \beta^{-\frac{\gamma}{2}}\mathbb{C}\left[  G_{0}%
,G_{0}\right]  \left(  \xi\right)  +2\beta^{-\frac{\gamma}{2}}\mathbb{C}%
\left[  G_{0},G_{1}\right]  \left(  \xi\right)  +2\beta^{-\frac{\gamma}{2}%
}\mathbb{C}\left[  G_{0},G_{2}\right]  \left(  \xi\right)  +\beta
^{-\frac{\gamma}{2}}\mathbb{C}\left[  G_{1},G_{1}\right]  \left(  \xi\right)
+\dots\right)
\end{align*}
where the operator $\mathbb{C}\left[  G_{0},G_{k}\right]  $ is the
bilinear operator defined in \eqref{eq:symbilcoll}.

Using that $\mathbb{C}\left[  G_{0},G_{0}\right]  =0$ as well as 
$\xi \cdot \partial_{\xi}G_{i}+3 G_{i}= \partial_{\xi}\cdot(\xi G_{i})$ for $i=0,1,2
$ we obtain:%
\begin{align}
&  \partial_{t}G_{1}+\partial_{t}G_{2}-\frac{\beta_{t}}{2\beta}\partial_{\xi
}\cdot(\xi G_{0})+\frac{\beta_{t}}{2\beta}\partial_{\xi}\cdot(\xi G_{1}%
)+\frac{\beta_{t}}{2\beta}\partial_{\xi}\cdot(\xi G_{2})+\ \label{H1E2}\\
&  -\partial_{\xi}\cdot\big(Q(t)\xi G_{0}\big)-\partial_{\xi}\cdot\big(Q(t)\xi G_{1}%
\big)-\partial_{\xi}\cdot\big(Q(t)\xi G_{2}\big)\nonumber\\
&  =\mu(t)\left( 2\beta^{-\frac{\gamma}{2}}\mathbb{C}\left[  G_{0}%
,G_{1}\right]  \left(  \xi\right)  +2\beta^{-\frac{\gamma}{2}}\mathbb{C}%
\left[  G_{0},G_{2}\right]  \left(  \xi\right)  +\beta^{-\frac{\gamma}{2}%
}\mathbb{C}\left[  G_{1},G_{1}\right]  \left(  \xi\right)  +...\right) .
\nonumber
\end{align}

We observe that we can expect to have a Hilbert like expansion only if the factors multiplying the collision terms are the largest. In particular, given that the largest terms on the left of \eqref{H1E2}, namely $\partial_{\xi}\cdot\big(Q(t)\xi G_{0}\big)$, we must require that $\mu(t) \beta^{-\frac{\gamma}{2}}\to \infty$ as $t\to \infty$. We will make this assumption in the following  and we will check at the end that this assumption is indeed satisfied. 
On the other hand, due to \eqref{eq:cdtHk}, we expect that the contributions of the terms $\frac{\beta_{t}}{2\beta}%
\xi\cdot\partial_{\xi}G_{1},\ \frac{\beta_{t}}{2\beta}\xi\cdot\partial_{\xi}G_{2}$ are
smaller than that of $\frac{\beta_{t}}{2\beta}\xi\cdot\partial_{\xi}G_{0}$ and
the contribution of $\partial_{\xi}\cdot\big(Q(t)\xi G_{2}\big)$ is smaller than that of $\partial_{\xi}\cdot\big(Q(t)\xi G_{1}\big).$ 
Analogously, $\partial_{t}G_{2}$
can be neglected compared with $\partial_{t}G_{1}.$ 
It is then natural to split \eqref{H1E2}  combining terms which can be expected to be of comparable order of magnitude.  
Notice that in the case 1) of the Conjecture we expect that $\frac{\beta_{t}}{2 \beta}$ is of order one (cf. \eqref{eq:beta1Hilb}).  Therefore, it would be natural to define the functions $G_1,\,G_2,\dots $ by means of the following hierarchy of equations: 
\begin{align}
& \label{H1E3}
 2 \mu(t) \beta^{-\frac{\gamma}{2}}\mathbb{C}\left[  G_{0}%
,G_{1}\right]  \left(  \xi\right) = 
\lambda_0(t)\partial_{\xi}\cdot(\xi
G_{0})-\partial_{\xi}\cdot\big(Q(t)\xi G_{0}\big) \\& 
 \label{H1E4}
2 \mu(t) \beta^{-\frac{\gamma}{2}}\mathbb{C}\left[  G_{0},G_{2}\right]  \left(  \xi\right) =
 \lambda_1(t)\partial_{\xi}\cdot(\xi
G_{0}) +\partial_{t} G_{1}+  \lambda_0(t)\partial_{\xi}\cdot(\xi G_{1})-\partial_{\xi}\cdot(Q(t)\xi G_{1})\\&
\hspace{4.2cm} -\mu
(t)\beta^{-\frac{\gamma}{2}}\mathbb{C}\left[  G_{1},G_{1}\right]  \left(
\xi\right)   \\& 
\qquad \dots\dots 
\nonumber 
\end{align}
where we used the decomposition 
\begin{equation}\label{eq:exbeta} 
\frac{\beta_{t}}{2\beta}=\lambda_0(t)+\lambda_1(t)+\lambda_2(t)+\dots .
\end{equation}
 The functions $\lambda_k(t)$ will be chosen in order to have suitable compatibility conditions for each equation of the hierarchy and can be expected to satisfy $\vert \lambda_0(t)\vert \gg \vert \lambda_1(t)\vert \gg \vert\lambda_2(t)\vert\gg\dots$ as $t\to\infty$.
\medskip

 On the other hand, in the case 2) of the Conjecture we expect  $\frac{\beta_{t}}{\beta}\to 0$ as $t\to\infty$ (cf., \eqref{eq:beta2Hilb}). Then, the hierarchy of equations yielding the functions
 $G_1,\,G_2,\dots $ is:
\begin{align}
\label{eq:secordHilb1}
2\mu(t)\beta^{-\frac{\gamma}{2}}\mathbb{C}\left[  G_{0},G_{1}\right]  \left(
\xi\right)   &  = -\partial_{\xi}\cdot (Q(t)\xi
G_{0})\\
2\mu(t)\beta^{-\frac{\gamma}{2}}\mathbb{C}\left[  G_{0},G_{2}\right]  \left(  \xi\right)   &  =
\left(  \partial_{t} G_{1}+ \frac{\beta_{t}%
}{2 \beta}\partial_{\xi}\cdot(\xi G_{0})-\partial_{\xi}\cdot(Q(t)\xi G_{1}) -\mu
(t)\beta^{-\frac{\gamma}{2}}\mathbb{C}\left[  G_{1},G_{1}\right]  \left(
\xi\right)  \right). \label{eq:secordHilb2}%
\end{align}

We will show now that the hierarchies \eqref{H1E3}-\eqref{H1E4} are consistent and \eqref{eq:secordHilb1}-\eqref{eq:secordHilb2} yield the asymptotics \eqref{eq:beta1Hilb}, \eqref{eq:beta2Hilb} respectively. 
\bigskip

We now rewrite the hierarchies \eqref{H1E3}-\eqref{H1E4} and \eqref{eq:secordHilb1}-\eqref{eq:secordHilb2}  in terms of the operator $\mathbb{L}$ defined in \eqref{eq:linL}. 
Using \eqref{eq:linL}, \eqref{equivGwH} as well as the energy conservation $\left\vert \xi^{\prime}\right\vert ^{2}+\left\vert \xi_{\ast}^{\prime}\right\vert ^{2}=\left\vert \xi\right\vert ^{2}+\left\vert \xi_{\ast}\right\vert ^{2}$we have:
\begin{align}\label{eq:CtoL}
\mathbb{C}\left[  G_{0},G_{k}\right]  \left(  \xi\right)  =\left(
C_{0}\right)  ^{2}\mathbb{C}\left[  e^{-\left\vert \xi\right\vert ^{2}%
},e^{-\left\vert \xi\right\vert ^{2}}H_{k}\right]  \left(  \xi\right) =\left(
C_{0}\right)  ^{2} \frac{e^{-\left\vert \xi\right\vert ^{2}}}{2}\mathbb{L}\left[
H_{k}\right]  \left(  \xi\right) .
\end{align}

We consider first the case 1) of the Conjecture. 
Using \eqref{eq:CtoL} we can rewrite \eqref{H1E3}, \eqref{H1E4} as 
\begin{align}
\label{eq:firstorderHilb}
\mu(t)\beta^{-\frac{\gamma}{2}} \left(
C_{0}\right)  ^{2} \mathbb{L}\left[
H_{k}\right]  \left(  \xi\right) & = J_k(\xi), \quad k=1,2,\dots 
\end{align}
where 
\begin{align}\label{eq:J1}
&J_1(\xi):= e^{\left\vert \xi\right\vert ^{2}}\left( \lambda_0(t)
\partial_{\xi}\cdot(\xi G_{0}) -\partial_{\xi}\cdot(Q(t)\xi G_{0})\right)
\\&
\label{eq:J2}
J_2(\xi):=e^{\left\vert \xi\right\vert ^{2}}\left( \lambda_1(t)\partial_{\xi}\cdot(\xi
G_{0}) +\partial_{t} G_{1}+  \lambda_0(t)\partial_{\xi}\cdot(\xi G_{1})-\partial_{\xi}\cdot(Q(t)\xi G_{1}) -\mu
(t)\beta^{-\frac{\gamma}{2}}\mathbb{C}\left[  G_{1},G_{1}\right]  \left(
\xi\right)  \right) \\&
\dots \nonumber
\end{align}
To solve \eqref{eq:firstorderHilb} for $k=1,2,\dots$ we have to impose  the Fredholm compatibility conditions 
\begin{equation}\label{eq:comp}
 \left\langle J_k , \Psi \right\rangle_w =0 \quad \text{for}\quad \Psi\in\{1,\xi,\vert\xi\vert^2\}.
\end{equation}
If $k=1$ the compatibility condition reduces to
\begin{align}\label{eq:compJ1}
 \left\langle J_1 , \Psi \right\rangle_w &=  
 \left\langle \partial_{\xi}\cdot \left(\left(  \lambda_0(t) \xi -Q(t)\xi \right)G_{0}\right), \Psi(\xi)\right\rangle \nonumber \\
 &=- \left\langle  \Big( \lambda_0(t) \xi  -Q(t)\xi \Big)G_{0}, \partial_{\xi}\Psi(\xi)\right\rangle=0,
\end{align}
where $\langle \cdot,\cdot\rangle$ denotes the usual scalar product in $L^2(\R^3)$. If $\Psi=1$ we have that \eqref{eq:compJ1} holds because $\partial_{\xi}\Psi(\xi)=0$. If 
$\Psi=\xi$ \eqref{eq:compJ1} follows as a consequence of the symmetry of $G_0$ under reflections. If $\Psi=\vert\xi\vert^2$, \eqref{eq:compJ1} becomes
\begin{align}\label{eq:compJ1}
 \lambda_0(t) \int\left\vert \xi\right\vert ^{2}\,G_{0}(\xi
)\,d\xi=\int(\xi\cdot Q(t)\xi) G_{0}(\xi)\,d\xi =\frac{1}{3}\Tr(Q(t))\int|\xi|^{2}G_{0}(\xi)\,d\xi . 
\end{align}
Therefore, using that $\frac{\beta_{t}}{2 \beta}\sim \lambda_0(t)$ and \eqref{eq:matrixQ} we obtain $\beta_{t}= \left(a+ O\left(\frac{1}{t^{1+\delta}}\right)\right)\beta$ as $t\to \infty$.  Hence, \eqref{eq:beta1Hilb} follows. 
Notice however that, in order to obtain a consistent expansion with the form \eqref{H1E1_bis} we need to have $\mu(t) \beta^{-\frac{\gamma}{2}}\to \infty$ as $t\to \infty$. Therefore, due to \eqref{eq:beta1Hilb}  we must impose the condition $\mu(t) e^{-\frac{\gamma}{2}at}\to \infty$ as $t\to \infty$, as requested in the case 1) of the Conjecture. Under this assumption we obtain iteratively the terms in the expansion \eqref{H1E1_bis}. More precisely, we have 
\begin{equation}\label{eq:H1}
H_1= \frac{\beta^{\frac{\gamma}{2}}}{C_{0}^{2}\mu(t)} \mathbb{L}^{-1}[J_1].
\end{equation}
 Plugging this into \eqref{eq:J2} we can solve \eqref{eq:firstorderHilb} for $k=2$ using the corrective term $\lambda_1(t)$ in \eqref{eq:exbeta} in order to obtain the compatibility condition for $J_2$. 
We observe that the compatibility condition 
\begin{equation}\label{eq:compCG1}
\left\langle \mathbb{C}\left[  G_{1},G_{1}\right], \Psi\right\rangle=0\quad \text{ for} \quad \Psi\in\{1,\xi,|\xi|^2\}
\end{equation} 
is automatically satisfied due to the properties of the collision operator, namely conservation of mass, momentum and kinetic energy. The contribution to the compatibility conditions due to the terms $\partial_{\xi}\cdot(\xi G_{1})$, $\partial_{\xi}\cdot(Q(t)\xi G_{1})$, $\partial_{t}G_{1}$ can be computed as in the case of the analogous terms in $J_1$.  The only non trivial compatibility condition turns out to be the one associated to $\Psi=\vert\xi\vert^2$ which implies that $\lambda_1(t)$ is of the same order of magnitude of $\vert G_1\vert.$  Due to \eqref{eq:H1} it follows that $\lambda_1(t)\leq C\, \frac{\beta^{\frac{\gamma}{2}}}{\mu(t)} \ll  \frac{1}{t^{1+\delta}}$ as $t\to\infty$.
Therefore $\lambda_1(t)$ does not modify the asymptotics \eqref{eq:beta1Hilb}.

\bigskip 

We now consider case 2) of the Conjecture. Using again \eqref{eq:CtoL} we can rewrite \eqref{eq:secordHilb1}, \eqref{eq:secordHilb2} as 
\begin{align}
\label{eq:Hilb_bis}
\mu(t)\beta^{-\frac{\gamma}{2}} \left(
C_{0}\right)  ^{2} \mathbb{L}\left[
H_{k}\right]  \left(  \xi\right) & = J_k(\xi), \quad k=1,2,\dots 
\end{align}
where now
\begin{align}\label{eq:J1bis}
&J_1(\xi):=- e^{\left\vert \xi\right\vert ^{2}} \partial_{\xi}\cdot(Q(t)\xi G_{0})
\\&
\label{eq:J2bis}
J_2(\xi):=e^{\left\vert \xi\right\vert ^{2}}\left(  \partial_{t} G_{1}+ \frac{\beta_{t}}{2 \beta}\partial_{\xi}\cdot(\xi G_{0})-\partial_{\xi}\cdot(Q(t)\xi G_{1}) -\mu(t)\beta^{-\frac{\gamma}{2}}
\mathbb{C}\left[  G_{1},G_{1}\right]  \left(\xi\right)  \right) \\&
\dots \nonumber
\end{align}

In this case, differently from case 1) the equation \eqref{eq:Hilb_bis} with $k=1$ does not provide any information about the function $\beta$. On the other hand, this equation for $k=1$ is always solvable because 
\begin{equation}\label{eq:compJ1bis}
 \left\langle J_1 , \Psi \right\rangle_w =0 \quad \text{for}\quad \Psi\in\{1,\xi,\vert\xi\vert^2\}.
\end{equation}
Indeed, in the case of $\Psi\in\{1,\xi \}$ this can be proved as we did in case 1). When $\Psi(\xi)=\vert\xi\vert^2$ we have
\begin{align}\label{eq:compJ1xi2}
 \left\langle J_1 , \vert\xi\vert^2 \right\rangle_w & = - \left\langle \partial_{\xi}\cdot(Q(t)\xi G_{0}), \vert\xi\vert^2 \right\rangle=2\left\langle Q(t)\xi G_{0} , \xi \right\rangle\nonumber\\&
=2 \int (\xi\cdot Q(t)\xi) G_{0}(\xi)\,d\xi =\frac{2}{3}\Tr(Q(t))\int|\xi|^{2}G_{0}(\xi)\,d\xi =0, 
\end{align}
due to \eqref{eq:case2Hilb}. 
Therefore, we obtain
\begin{equation}
\label{eq:G1_bis}
H_{1}(\xi)=-\frac{\beta^{\frac{\gamma}{2}}}{C_0^2\mu
(t)}\,\mathbb{L}^{-1}[e^{|\xi|^{2}} \partial_{\xi}\cdot (Q(t)\xi G_{0})] .
\end{equation}

We will use repeatedly in the following computations the identity
\[
\left\langle e^{\vert\xi\vert^2}f
,g\right\rangle _{w} =\left\langle f ,g \right\rangle .
\] 

We can now obtain a differential equation for $\beta$ imposing the Fredholm compatibility conditions in \eqref{eq:Hilb_bis} for $k=2$, namely
\begin{equation}\label{eq:compJ2bis}
 \left\langle J_2 , \Psi \right\rangle_w =0 \quad \text{for}\quad \Psi\in\{1,\xi,\vert\xi\vert^2\}.
\end{equation}
The compatibility condition for $\Psi\in\{1,\xi \}$ follows from the fact 
that 
\begin{align}
\label{eq:cmpcdt_1} & 
\left\langle G_1,\Psi \right\rangle =\left\langle \partial_{\xi}\cdot (\xi G_{0})(\xi),\Psi \right\rangle = \left\langle \partial_{\xi}\cdot (Q(t) \xi G_{1})(\xi),\Psi \right\rangle
=0 \quad \text{for}\quad \Psi\in\{1,\xi \},
\end{align}
as well as \eqref{eq:compCG1}. Then, the only nontrivial compatibility condition is the one for $\Psi=\vert\xi\vert^2$. Using that $\left\langle G_1, \vert\xi\vert^2\right\rangle =0$ and 
\eqref{eq:compCG1}, this compatibility condition reduces to
\[
\frac{\beta_{t}}{2\beta}\left\langle \partial_{\xi}\cdot (\xi G_{0})(\xi),|\xi
|^{2}\right\rangle =\left\langle \partial_{\xi}\cdot (Q(t) \xi G_{1})(\xi),|\xi
|^{2}\right\rangle ,
\]
or, equivalently,
\[
\frac{\beta_{t}}{2\beta}\left\langle \xi G_{0}(\xi), \xi\right\rangle
=\left\langle Q(t) \xi G_{1}(\xi),\xi\right\rangle .
\]
Therefore, using \eqref{equivGwH} and \eqref{eq:G1_bis}, we obtain
\begin{equation}\label{eq:betasecord}
\frac{\beta_{t}}{2\beta}=\frac{\left\langle \xi, Q(t) \xi G_{1}(\xi)
\right\rangle }{\int\xi^{2} G_{0}(\xi)d\xi} = \frac{\beta^{\frac{\gamma
}{2}}}{C_{0}\mu(t)} \frac{\left\langle \xi, Q(t) \xi\, e^{-|\xi|^{2}}(-\mathbb{L})^{-1}[e^{|\xi|^{2}}\partial_{\xi}\cdot(Q(t)\xi G_{0})] ) \right\rangle }{\int\xi^{2}
G_{0}(\xi)d\xi}.
\end{equation}

We observe that 
$$
\partial_{\xi}\cdot(Q(t)\xi G_{0})=
\partial_{\xi}\cdot (Q(t)\xi)G_{0} -2 (\xi\cdot Q(t)\xi) G_{0}(\xi)=-2 (\xi\cdot Q(t)\xi) G_{0}(\xi)
$$
since $\partial_{\xi}\cdot(Q(t)\xi)=\Tr(Q(t))=0$. Then, \eqref{eq:betasecord} becomes
\begin{align*}
\frac{\beta_{t}}{2\beta} & =\frac{2}{C_{0}} \frac{\beta^{\frac{\gamma}{2}}}{\mu(t)}
\frac{\left\langle \xi, Q(t) \xi\, e^{-|\xi|^{2}}(\mathbb{L})^{-1}[e^{|\xi|^{2}} (\xi\cdot
Q(t)\xi)  G_{0}(\xi)] ) \right\rangle }{\int\xi^{2} G_{0}(\xi)d\xi
}\\&
=2 \frac{\beta^{\frac{\gamma}{2}}}{ \mu(t)} \frac{\left\langle \xi\cdot Q(t)
\xi, (\mathbb{L})^{-1}[ (\xi\cdot Q(t)\xi)] ) \right\rangle _{w} }{\int\xi^{2} G_{0}(\xi)d\xi} \\&
=\frac{2}{C_0} \frac{\beta^{\frac{\gamma}{2}}}{ \mu(t)} \frac{\left\langle \xi\cdot Q(t)
\xi, (\mathbb{L})^{-1}[ (\xi\cdot Q(t)\xi)] ) \right\rangle _{w} }{\int\xi^{2} e^{-\vert\xi\vert^2}d\xi}\\&
=\frac{4}{3} \frac{\beta^{\frac{\gamma}{2}}}{ \mu(t)}
\left\langle \xi\cdot Q(t) \xi, (\mathbb{L})^{-1}[ (\xi\cdot Q(t)\xi)] ) \right\rangle _{w}  =:-\frac{4}{3}b (1+o(1))\frac{\beta^{\frac{\gamma}{2}}}{ \mu(t)},
\end{align*}
where we used \eqref{eq:defC0} and $b>0$ is as in \eqref{eq:GreenKubo}. 
Then, we get
\[
\beta_{t}=-\frac{8}{3}b (1+o(1)) \frac{\beta^{\frac{\gamma}{2}+1}}{\mu(t)}.
\]
Solving this differential equation, using the definition of $\lambda$, we obtain
\[
\beta(t) \sim \left( \frac{4}{3} \gamma\, b \lambda(t) \right)^{-\frac{2}{\gamma}}(1+o(1)) \quad \text{as}\quad t\to\infty.
\]
Notice that in order to obtain an expansion of the form \eqref{H1E1_bis} we used, as in the case 1) of the Conjecture, that $\mu(t)\beta^{-\frac{\gamma}{2}}\to\infty$. Due to \eqref{eq:beta2Hilb} this follows if $\mu(t)\lambda(t)\to\infty$ as $\lambda(t)\to\infty$. The definition of $\lambda(t)$ implies that this is equivalent to $\frac{\lambda(t)}{\lambda'(t)}\to \infty$ which follows from the assumption in the statement of the Conjecture.
\end{proofof}

\begin{remark}
Notice that in the both cases of the Conjecture the class of solutions described has an  increasing temperature as $t\to\infty$ if $\gamma>0$ and a decreasing temperature if $\gamma<0$.  
This might be physically expected because the collision operator yields a larger contribution if the temperature of the distribution is higher and $\gamma>0$ or  the temperature of the distribution is lower and $\gamma<0$. 
\end{remark}

\subsection{Applications}
\label{sec:HilbExp}

We now describe the consequences of the Conjecture \ref{th:genHilbexp} for the long time asymptotics of homoenergetic flows (cf., \eqref{S8E9}) for different choices of the matrix $L(t)$ according to the classification given in Theorem \ref{ClassHomEne}.

\subsubsection{Maxwellian distribution as long time asymptotics for simple
shear and kernels with homogeneity $\gamma>0$}
\label{ss:Hilbsimpleshear}

In this subsection we study the long time asymptotics of solutions of
\begin{align}
\label{eq:BoltzshearHilb1}\partial_{t}g-Kw_{2}\partial_{w_{1}}g  &  =
\int_{\mathbb{R}^{3}}dw_{\ast} \int_{S^{2}} d\omega\, B\left(  n\cdot \omega,\left\vert w-w_{\ast}\right\vert \right)  \left[
g^{\prime}g_{\ast}^{\prime}-g_{\ast}g\right]
\end{align}
when the cross-section of the collision kernel has homogeneity larger than zero. This corresponds to take the matrix $L(t)$ of the form \eqref{T1E5} in \eqref{S8E9}.

More precisely, we will assume for definiteness that the kernels $B\left(
n\cdot \omega,\left\vert v-v_{\ast}\right\vert \right)  $ are homogeneous in
$\left\vert v-v_{\ast}\right\vert $ with homogeneity $\gamma>0$, cf.,~\eqref{S8E7}.

Moreover, we assume also that $B$ is bounded if $\left\vert v-v_{\ast
}\right\vert \leq1,$ i.e. there is nonsingular behavior in the angular
variable $\omega.$ Strictly speaking this is not needed, and this condition
could be replaced by a more general condition yielding convenient functional analysis properties for the operator $\mathbb{L}$ defined in \eqref{eq:linL}. 
A typical example of cross-section is the
one of hard-sphere potentials,
\[
B\left(  \omega,\left\vert v-v_{\ast}\right\vert \right)  =\left\vert
\omega\cdot\left(  v-v_{\ast}\right)  \right\vert \ \ ,\ \ e=\omega
\]

The intuitive idea behind the asymptotic behavior computed in this case 
is that there is a competition between the
shear term $w_{2}\partial_{w_{1}}g$ and the collision term $\mathbb{C}g\left(
w\right)  $ for large $\left\vert w\right\vert$.  The effect of the shear term is to increase the temperature of the system.  Comparing the order of magnitude of the shear term and the collision term, it turns out that for $\gamma>0$,  the collision
term becomes the dominant one for large times. Therefore, the distribution of particles approaches a Maxwellian distribution. The effect of the shear is a small perturbation, compared to the collision term which produces a growth of the temperature  of the Maxwellian. Since in this case $\Tr(L(t))=0$ the asymptotics of the solution of \eqref{eq:BoltzshearHilb1} can be computed using the case 2)
of the Conjecture \ref{th:genHilbexp}.
More precisely, the main
result of this subsection is the following Conjecture.

\begin{conjecture}
\label{th:MdShear} 
Suppose that the cross-section
$B(\cdot,\cdot)$ satisfies condition \eqref{S8E7} with $\gamma>0$. Then there exists  $g\left(  t,w\right)  $ a weak solution of the
Boltzmann equation \eqref{eq:BoltzshearHilb1} in the sense of Definition \ref{WeakSol} for which the following asymptotics
as $t \to\infty$ holds:
\[
\label{eq:MaxwHilb}\beta(t)^{-\frac{3}{2}}g\big(t,\frac{\xi}{\sqrt{\beta}%
}\big) \to C_{0} e^{-\left\vert \xi\right\vert ^{2}} \quad\text{in}\quad
L^{2}\left(  \mathbb{R}^{3};e^{-\left\vert \xi\right\vert ^{2}}d\xi\right) ,
\]
with $C_0=\pi^{-\frac{3}{2}}$. 
Moreover, $\beta\left(  t\right)  $ satisfies
\[
\beta\left(  t\right) ={C}\,{t^{-\frac{2}{\gamma}}}(1+o(1))\ \ \text{as\ \ }%
t\rightarrow\infty.
\]
\end{conjecture}

To support the conjecture above we resort to the general strategy proposed in Section
\ref{sec:genHilbexp}. Indeed, we can apply Conjecture \ref{th:genHilbexp}, case
2) to \eqref{eq:BoltzshearHilb1}. Note that in this case $\mu(t)=1$,
$\gamma>0$ and $\Tr(L)=0$.
Thus, since $\lambda(t)=t$, we have that
\[
\beta\left(  t\right)  =
C t^{-\frac{2}{\gamma}}(1+o(1))\ \ \text{as\ \ }t\rightarrow\infty,
\]
where $C=\left(\frac{4}{3} \gamma\, b\right)^{-\frac{2}{\gamma}}$ and $b>0$ is given by:
\[
b=K^2\left\langle \xi_{1}\xi_{2},\left(
-\mathbb{L}\right)  ^{-1}\left(  \xi_{1}\xi_{2}\right)  \right\rangle .
\]

Therefore, $\beta\rightarrow0$ as $t\to\infty$ as expected and the temperature
increases as a power law. We notice that the exponent is larger if $\gamma$
approaches zero and if $\gamma=0$ we have the exponential growth described in
\cite{JNV1}, Subsection 5.1.1.


\subsubsection{Maxwellian distribution as long time asymptotics for planar
shear with $K=0$ and kernels with homogeneity $\gamma<0$}

\label{ss:Hilb1dil}

We compute here asymptotic formulas for the long time asymptotics of
homoenergetic solutions of
\begin{equation}
\label{eq:Boltz1dilHilb1}\partial_{t}g-\frac{1}{t}w_{3}\partial_{w_{3}}%
g=\int_{\mathbb{R}^{3}}dw_{\ast} \int_{S^{2}}d\omega\, B\left(  \omega
,\left\vert w-w_{\ast}\right\vert \right)  \left[  g^{\prime}g_{\ast}^{\prime
}-g_{\ast}g\right]
\end{equation}
with cross-sections $B$ satisfying \eqref{S8E7} with $\gamma<0$.  
Notice that in this case we are choosing $L(t)$ as in \eqref{T1E3} in the homoenergetic flows \eqref{S8E9}. 
In this case the dominant term for large velocities is the collision term and
we can obtain this asymptotics using the Hilbert expansion approach proposed
in Subsection \ref{sec:genHilbexp}.
Since $\Tr(L(t))\neq 0$ we can reduce the problem to the case 1) of Conjecture \ref{th:genHilbexp}. Given that $\gamma<0$ the solutions behave like  Maxwellians for long times and have a decreasing temperature. More precisely, the main result of this subsection is
the following Conjecture.

\begin{conjecture}
\label{th:Md1dil} Suppose that the cross-section
$B(\cdot,\cdot)$ satisfies condition \eqref{S8E7} with $\gamma<0$.
Then there exists  $g\left(  t,w\right)  $ a weak solution of the
Boltzmann equation \eqref{eq:Boltz1dilHilb1} in the sense of Definition \ref{WeakSol} for which the following asymptotics
as $t \to\infty$, holds:
\[
\label{eq:MaxwHilb}g\big(t,\frac{\xi}{\sqrt{\beta}}\big)\to C_{0}%
e^{-\left\vert \xi\right\vert ^{2}}\quad\text{in}\quad L^{2}\left(
\mathbb{R}^{3};e^{-\left\vert \xi\right\vert ^{2}}d\xi\right) 
\]
for some $C_0>0$. 
Moreover, $\beta\left(  t\right)  $ satisfies
\[
\beta\left(  t\right)  ={C}\,{t^{\frac2 3}}(1+o(1))\ \ \text{as\ \ }t\rightarrow
\infty.
\]
\end{conjecture}

\medskip
We notice that Conjecture \ref{th:Md1dil} is supported by the general
strategy proposed in Subsection \ref{sec:genHilbexp} using the explicit
expression for $L$ given by \eqref{T1E3} with $K=0$. More precisely, we
rescale the solution $g$ as $g(t,w)=\frac{1}{t}\bar{g}(t,w)$ which solves
\[
-\frac{1}{t^{2}}\bar{g}+\frac{1}{t}\partial_{t}\bar{g}-\frac{1}{t^{2}}
w_{3}\partial_{w_{3}}\bar{g}=\frac{1}{t^{2}} \mathbb{C}\bar{g}%
\]
and perform the change of variables $\tau=\log t$ so that we obtain
\[
\partial_{\tau}\bar{g}-\partial_{w_{3}}\left(  w_{3}\bar{g}\right)
=\mathbb{C}\bar{g} .
\]
We can now apply Conjecture \ref{th:genHilbexp}, case 1) to the equation
above. Note that here $\mu(\tau)=1$, $a=\frac2 3 $ since $\Tr(L)=1$ and
$e^{-\frac{\gamma}{2}a\tau}=e^{\frac{|\gamma|}{3}\tau}\to\infty$ as $\tau
\to\infty$. 
Therefore $\beta(\tau)=Ce^{\frac{2}{3}\tau}(1+o(1))$ as $\tau\to\infty$
which implies that $\beta(t)=Ct^{\frac2 3}(1+o(1))$ as $t\to\infty$.

\subsubsection{Maxwellian distribution as long time asymptotics for
cylindrical dilatation and cylindrical dilatation with shear for kernels with homogeneity $\gamma<\gamma_{crit}$}

\label{ss:Hilb2dil}

We are interested here in the long time asymptotics of
\begin{equation}
\partial_{t}g-\frac{1}{t}\left(  (w_{1}+Kw_3)\partial_{w_{1}}+w_{2}\partial_{w_{2}
}\right)  g=\int_{\mathbb{R}^{3}}dw_{\ast}\int_{S^{2}}d\omega\,B\left(
\omega,\left\vert w-w_{\ast}\right\vert \right)  \left[  g^{\prime}g_{\ast
}^{\prime}-g_{\ast}g\right]  \label{eq:Boltz2dilHilb1}%
\end{equation}
with cross-section $B$ satisfying \eqref{S8E7}. In this case we choose $L(t)$ as in \eqref{T1E2} in \eqref{S8E9}. It is possible to obtain solutions behaving asymptotically as Maxwellians for long times if the homogeneity $\gamma<\gamma_{crit}$ with $\gamma_{crit}:=-\frac{3}{2}$. 

Also in this case, given that $\Tr(L(t))\neq 0$, we will be able to reduce the problem to the case 1)  in Conjecture \ref{th:genHilbexp}. Moreover, since $\gamma<0$ the temperature of the Maxwellian is decreasing.
The main result of this subsection is the following Conjecture.

\begin{conjecture}
\label{th:Md2dil} Suppose that the cross-section
$B(\cdot,\cdot)$ satisfies condition \eqref{S8E7} with $\gamma< \gamma_{crit}=-\frac{3}{2}$. Then there exists  a weak solution 
$g\left(  t,w\right)  $ of the
Boltzmann equation \eqref{eq:Boltz2dilHilb1} in the sense of Definition \ref{WeakSol} for which the following asymptotics
as $t \to\infty$, holds:
\[
\label{eq:MaxwHilb}g\big(t,\frac{\xi}{\sqrt{\beta}}\big)\to C_{0}%
e^{-\left\vert \xi\right\vert ^{2}}\quad\text{in}\quad L^{2}\left(
\mathbb{R}^{3};e^{-\left\vert \xi\right\vert ^{2}}d\xi\right)  .
\]
for some $C_0>0$.  
Moreover, $\beta\left(  t\right)  $ satisfies
\[
\beta\left(  t\right)  ={C}\,{t^{\frac4 3}}\ \ \text{as\ \ }t\rightarrow
\infty.
\] 
\end{conjecture}


\medskip 
Conjecture \ref{th:Md2dil} is supported following the general
strategy proposed in Subsection \ref{sec:genHilbexp}. 
Since the density $\int_{{\mathbb{R}}^{3}}g(t,w)dw$ behaves for large $t$ as
$\frac{1}{t^{2}}$,
it is convenient to renormalize the solution $G$ setting $g(t,w)=\frac
{1}{t^{2}}\bar{g}(t,w)$.
Rescaling also the time variable $t$ so that $\log(t)=\tau$ we get
\begin{equation}
\label{eq:2ddileq}\partial_{\tau}\bar{g}-\partial_{w}\cdot\left(  Lw \bar
{g}\right)  =e^{-\tau} \mathbb{C}G.
\end{equation}
Observe that $\mu(\tau)=e^{-\tau}$, $a= \frac{4}{3}$. Therefore
\[
\mu(\tau)e^{-\frac{\gamma}{2}a\tau}=e^{-(1+\frac{2}{3}\gamma)\tau}\to\infty
\]
for $\gamma<-\frac{3}{2}<0$, which gives the critical threshold for the homogeneity
$\gamma$. 

We then apply Conjecture \ref{th:genHilbexp}, case 1) to \eqref{eq:2ddileq}
to obtain 
\[
\beta^{-\frac3 2}\bar{g}\left( t,\frac{\xi}{\sqrt{\beta}}\right) \to C_{0}
e^{-|\xi|^{2}},
\]
with $\beta(\tau)=Ce^{\frac4 3\tau}$ as $\tau\to\infty$. Coming back to the
original variable $t$, we get
\[
g\left( t,\frac{\xi}{\sqrt{\beta}}\right) \to C_{0} e^{-|\xi|^{2}},
\]
with $\beta(t)=Ct^{\frac4 3}(1+o(1))$ as $t\to\infty$.

\subsubsection{Maxwellian distribution as long time asymptotics for
homogeneous dilatation and kernels with homogeneity $\gamma<-2$}

\label{sss:Hilb3dil}

In this subsection we consider the case of homogeneous dilatation. This case has been
considered in detail in \cite{Nikol1}, \cite{Nikol2}. In those papers, it was shown that homoenergetic solutions in the case of homogeneous dilatation can be reduced, by means of a suitable change of variables, to the analysis of the homogenous Boltzmann equation whose mathematical theory is very well developed (cf.,~for instance \cite{Car},~\cite{CIP},~\cite{MW},~\cite{V02}). For the sake of completeness, we briefly recall the results
obtained in those papers. In
\cite{Nikol1}, \cite{Nikol2} it has been proved that for interaction
potentials with the form $V\left(  x\right)  =\frac{1}{\left\vert x\right\vert
^{\nu-1}}$ the solution of the Boltzmann equation in the case
of homogeneous dilatation does not approach to the Maxwellian if $\nu>\frac{7}{3}$ and it does if
$\nu\leq\frac{7}{3}$.  
We recall that the homogeneity of the collision kernel is
$\gamma=\frac{\nu-5}{\nu-1}$ and Maxwellian molecules correspond to $\nu=5$. 
Moreover, the case of
homogeneous dilatation has been discussed also in \cite{TM}. According to
\cite{TM}, convergence to the Maxwellian in the case of
homogeneous dilatation fails for Maxwellian molecules (i.e.,~$\gamma=0$), but
but holds in the case of condensation near the blow-up time for the density.

We consider homoenergetic flows (\ref{B1_0}) with $L(t)$ given by
\eqref{T1E1}. Then $g=g\left(  t,w\right)  $ satisfies%
\begin{equation}
\partial_{t}g-\left[  \frac{I}{t}+\alpha\left(  t\right)  \right]
w\cdot\partial_{w}g=\mathbb{C}g\left(  w\right)  \label{T6E5}%
\end{equation}
where $\alpha\left(  t\right)  $ is a matrix satisfying $\alpha\left(
t\right)  =O\left(  \frac{1}{t^{2}}\right)  $ as $t\rightarrow\infty.$

Using (\ref{S8E5}) we obtain
\[
\rho\left(  t\right)  =\int_{\mathbb{R}^{3}}g\left(  t,dw\right)  =\frac
{\rho\left(  1\right)  }{t^{3}}\exp\left(  -\int_{1}^{t}\operatorname*{tr}%
\alpha\left(  s\right)  ds\right)  \ \ \text{as\ \ }t\rightarrow\infty.
\]

We define%
\begin{equation}
\bar{g}\left(  t,w\right)  =t^{3}\exp\left(  \int_{1}^{t}\operatorname*{tr}%
\alpha\left(  s\right)  ds\right)  g\left(  t,w\right)  . \label{T6E6b}%
\end{equation}
Then
\begin{equation}
\partial_{t}\bar{g}-\partial_{w}\cdot\left(  \left[  \frac{w}{t}+\alpha\left(
t\right)  w\right]  \bar{g}\right)  =\mathbb{C}\bar{g}\left(  w\right)
\label{T6E6}%
\end{equation}
and $\partial_{t}\left(  \int_{\mathbb{R}^{3}}\bar{g}\left(  t,dw\right)
\right)  =0.$

On the other hand, it is easy to derive a formula for the average internal
energy $\bar{\varepsilon}\left(  t\right)  =\int_{\mathbb{R}^{3}}\left\vert
w\right\vert ^{2}\bar{g}\left(  t,dw\right)  $ in these flows. Indeed, we have:%
\[
\partial_{t}\bar{\varepsilon}\left(  t\right)  +\left[  \frac{2}{t}+O\left(
\frac{1}{t^{2}}\right)  \right]  \bar{\varepsilon}\left(  t\right)  =0
\]
whence $\bar{\varepsilon}\left(  t\right)  =\frac{C}{t^{2}}\left(  1+O\left(
\frac{1}{t}\right)  \right)  $ as $t\rightarrow\infty.$ Therefore, the average
velocity of the particles tends to zero as $t\rightarrow\infty$ and it behaves
as $\frac{1}{t}.$ This suggests we look for solutions of the following form:
\begin{equation}
\bar{g}\left(  t,w\right)  =t^{3}G\left( \tau, \xi\right)  \ \ ,\ \ \xi
=wt,\ \ \tau=\log\left(  t\right)  . \label{T6E6a}%
\end{equation}

Using the fact that the collision kernel $B$ is homogeneous of order $\gamma$
we obtain%
\begin{equation}
\partial_{\tau}G-\partial_{\xi}\cdot\left(  \bar{\alpha}\left(  \tau\right)
\xi G\right)  =e^{-\left(  2+\gamma\right)  \tau}\mathbb{C}G\left(
\xi\right)  \label{T6E7}%
\end{equation}
where $\left\vert \bar{\alpha}\left(  \tau\right)  \right\vert \leq Ce^{-\tau
}.$ It is then natural to introduce a new time scale
\[
ds=e^{-\left(  2+\gamma\right)  \tau}d\tau.\
\]

We note that we could have two possibilities: $\gamma<-2$ or $\gamma>-2$. Here
we consider the case $\gamma<-2$ while the case $\gamma>-2$ will be discussed
in \cite{JNV3}. Thus, $\left(  2+\gamma\right)  <0$ and
\[
s=\frac{e^{\left\vert 2+\gamma\right\vert \tau}-1}{\left\vert 2+\gamma
\right\vert }%
\]
whence $s\rightarrow\infty$ as $\tau\rightarrow\infty.$ We then have
\begin{equation}
\label{eq:Boltz3dilHilb1}\partial_{s}G-\partial_{\xi}\cdot\left(
\kappa\left(  s\right)  \xi G\right)  =\mathbb{C}G\left(  \xi\right)
\end{equation}
where $\left\vert \kappa\left(  s\right)  \right\vert \leq Ce^{-\left(
1+\left\vert 2+\gamma\right\vert \right)  \tau}\leq\frac{C}{s^{1+\frac
{1}{\left\vert 2+\gamma\right\vert }}}$ as $s\rightarrow\infty.$ Therefore, as $s\rightarrow\infty$, $G$
converges to a Maxwellian distribution with mass one
and energy of order one.

More precisely we have the following
\begin{conjecture}
\label{th:Md3dil} Suppose that the cross-section
$B(\cdot,\cdot)$ satisfies condition \eqref{S8E7} with $\gamma<-2$. Then there exists   a weak solution $g\left(  t,w\right)  $ of the
Boltzmann equation \eqref{eq:Boltz3dilHilb1} in the sense of Definition \ref{WeakSol} for which the following asymptotics
as $t \to\infty$, holds:
\[
\label{eq:MaxwHilb}g\big(t,\frac{\xi}{\sqrt{\beta}}\big) \to C_{0}
e^{-\left\vert \xi\right\vert ^{2}} \quad\text{in}\quad L^{2}\left(
\mathbb{R}^{3};e^{-\left\vert \xi\right\vert ^{2}}d\xi\right)  .
\]
for some $C_0>0$. Moreover, $\beta\left(  t\right)  $ satisfies
\[
\beta\left(  t\right)  ={C}\,{t^{2}}(1+o(1))\ \ \text{as\ \ }t\rightarrow\infty.
\]
\end{conjecture}
\medskip

We remark that a similar result holds when $\gamma=-2$ although in this case
the right time scale is $s=\tau.$

\bigskip

\subsubsection{Maxwellian distribution as long time asymptotics for combined orthogonal
shears \label{MaxComSOD} and $\gamma>0$.}

\bigskip

We now consider the long time asymptotics of homoenergetic flows (\ref{B1_0})
with $L\left(  t\right)  $ as in (\ref{T1E4}). Then $g$ solves
\begin{equation}
\partial_{t}g-\left[  K_{3}w_{2}+\left(  K_{2}-tK_{1}K_{3}\right)
w_{3}\right]  \partial_{w_{1}}g-K_{1}w_{3}\partial_{w_{2}}g=\mathbb{C}g\left(
w\right)  \label{eq:BoltzcsHilb1}%
\end{equation}
with cross-sections $B$ satisfying \eqref{S8E7} with $ \gamma>0$.

We first notice that from \eqref{eq:BoltzcsHilb1} we get $\partial_{t}\left(
\int_{\mathbb{R}^{3}}g\left(  t,dw\right)  \right)  =0.$ 
 Furthermore, since the
homogeneity $\gamma>0$ we expect to have solutions of
\eqref{eq:BoltzcsHilb1} in the form of Maxwellians
with increasing temperature. Given that $\Tr(L(t))=0$ we would be able to reduce the problem to one of the equations that can be studied  as in case 2) of Conjecture \ref{th:genHilbexp}.
Indeed, the main result of this subsection is the following Conjecture.

\begin{conjecture}
\label{th:MdCmShear}
Suppose that the cross-section
$B(\cdot,\cdot)$ satisfies condition \eqref{S8E7} with $\gamma>0$. Then there exists  a weak solution $g\left( t,w\right) $ of the
Boltzmann equation \eqref{eq:BoltzcsHilb1} in the sense of Definition \ref{WeakSol} for which the following asymptotics
as $t \to\infty$, holds:
\[
\beta(t)^{-\frac{3}{2}}g\big(t,\frac{w}{\sqrt{\beta}}\big) \to C_{0}
e^{-\left\vert w\right\vert ^{2}} \quad\text{in}\quad L^{2}\left(
\mathbb{R}^{3};e^{-\left\vert w\right\vert ^{2}}dw\right)  .
\]
with $C_0=\pi^{-\frac{3}{2}}$. 
Moreover, $\beta\left(  t\right)  $ satisfies
\[
\beta\left(  t\right)  = {C}\,{t^{-\frac{6}{\gamma}}}(1+o(1)) \ \ \text{as\ \ }%
t\rightarrow\infty.
\] 
\end{conjecture}
\medskip

In order to justify the conjecture above we change variables setting $\tau
=\frac{ t^{2}}{ 2}$ so that \eqref{eq:BoltzcsHilb1} becomes 
\begin{equation}
\partial_{\tau}g- \partial_{w_{1}}\left(  \left(  \frac{1}{\sqrt{2\tau}}
(K_{3}w_{2}+ K_{2}w_{3})- K_{1}K_{3}w_{3}\right)  g\right)  -\partial_{w_{2}%
}\left( \frac{1}{\sqrt{2\tau}}K_{1}w_{3}g\right) =\frac{1}{\sqrt{2\tau}%
}\mathbb{C}g\left(  w\right) .\label{eq:BoltzcsHilb1bis}%
\end{equation}
Since $\Tr(Q(\tau))=0$, $\mu(\tau
)=\frac{1 }{\sqrt{2\tau}}$ we can then apply Conjecture \ref{th:genHilbexp},
case 2) to the equation above and we obtain that $\beta(\tau)=C\tau^{-\frac
{3}{\gamma}}(1+o(1))$ for $\tau\to\infty$ 
where 
$C=\left(\frac{4}{3} \gamma\, b\right)^{-\frac{2}{\gamma}}$ and $b>0$ is given by:
\[
b=(K_1K_3)^2\left\langle \xi_{1}\xi_{3},\left(
-\mathbb{L}\right)  ^{-1}\left(  \xi_{1}\xi_{3}\right)  \right\rangle .
\]
Coming back to the original variable $t$,
we obtain that $\beta(t)=2^{\frac{3}{\gamma}}Ct^{-\frac{6}{\gamma}}(1+o(1))$ as $t\to\infty$. 

\bigskip

\subsubsection{Maxwellian distribution as long time asymptotics for simple shear with decaying planar dilatation/shear and $\gamma>0$} \label{ssec:sheardil}

We consider here the case of homoenergetic flows with $L(t)$ as in \eqref{T1E6}. More precisely we look at the asymptotics of the solution of the equation
\begin{align}
\label{eq:sheardecaying}
\partial_{t}g-K_2 w_2\partial_{w_1}g-\frac{1}{t}\partial_{w}\cdot\big(L_1wg\big)  &  =\mathbb{C}g\left(
w\right) 
\end{align}
where we have written \eqref{T1E6} as 
\begin{equation}
L(t)=L_0+\frac 1 t L_1+O\left(\frac{1}{t^2}\right)
\end{equation}
with $\Tr(L_0)=0$ and $\Tr(L_1)=1$. Note that in \eqref{eq:sheardecaying} we neglected the contribution of the integrable term $O\left(\frac{1}{t^2}\right)$.

In spite of the fact that $\Tr(L(t))$ vanishes at the leading order we cannot reduce this case to the case 2) of Conjecture \ref{th:genHilbexp} since for the resulting matrix $Q(t)$ we have $\Tr(Q(t))\neq 0$. Nevertheless, we can obtain the long time asymptotics of the corresponding homoenergetic solution adapting the ideas used to justify Conjecture \ref{th:genHilbexp}, case 2).
We observe that, since $\Tr(L(t))\sim \frac{1}{t}$ we have that the density $\rho(t)$ decays as $\frac{C}{t}$ as $t\to\infty$. Then, it is convenient to change variables and set  
$g(w,t)=\frac{1}{t}\bar{g}(w,t)$. Then $\bar{g}$ solves
\begin{align}
\label{eq:sheardecaying_22}
\partial_{t}\bar{g}-K_2 w_2\partial_{w_1}\bar{g}-\frac{1}{t}\partial_{w}\cdot\big(L_1w\bar{g}\big)  =\frac{1}{t}\mathbb{C}\bar{g}\left(w\right). 
\end{align}
We now introduce the function $G$ as in \eqref{eq:ansHilbgen} with $g$ replaced by $\bar{g}$. Therefore, $G$ satisfies \eqref{eq:ansHilbgen_1} with $Q(t)=L_0+\frac{1}{t}L_1$ and $\mu(t)=\frac{1}{t}$. 
We then expand $G$ as in \eqref{H1E1_bis} and, arguing as in the justification of the case 2) of Conjecture \ref{th:genHilbexp}, we obtain that the functions $H_k$ satisfy \eqref{eq:Hilb_bis} where the functions $J_k$ are now given by 
\begin{align}\label{eq:J1bis_sheardec}
&J_1(\xi):= - e^{\left\vert \xi\right\vert ^{2}}K_2 \partial_{\xi_1}\cdot(\xi_2 G_{0})
\\&
\label{eq:J2bis_sheardec}
J_2(\xi):=e^{\left\vert \xi\right\vert ^{2}}\left(  \partial_{t} G_{1}+ \frac{\beta_{t}}{2 \beta}\partial_{\xi}\cdot(\xi G_{0})-K_2 \partial_{\xi_1}\cdot(\xi_2 G_{1}) -\frac{1}{t}\partial_{\xi}\cdot\big(L_1\xi G_0\big)  -\mu(t)\beta^{-\frac{\gamma}{2}}
\mathbb{C}\left[  G_{1},G_{1}\right]  \left(\xi\right)  \right) \\&
\dots \nonumber
\end{align}
Notice that this expansion is consistent as long as $\mu(t)\beta^{-\frac{\gamma}{2}}\to \infty$ as $t\to \infty$. The function $H_1$ is now given by 
\begin{equation}
\label{eq:G1_bis_sheardec}
H_{1}(\xi)=-\frac{\beta^{\frac{\gamma}{2}}}{C_0^2\mu
(t)}\,\mathbb{L}^{-1}[e^{|\xi|^{2}}K_2 \partial_{\xi_1}\cdot(\xi_2 G_{0}) ] .
\end{equation}
On the other hand, the compatibility condition \eqref{eq:Hilb_bis} for $k=2$ yields a differential equation for $\beta$, namely 
\begin{align*}
\frac{\beta_{t}}{2\beta} 
=-\frac{4b}{3}\frac{\beta^{\frac{\gamma}{2}}}{ \mu(t)} +\frac{1}{t}\frac{\left\langle L_1 \xi G_0,\xi \right\rangle}{\int\xi^{2} G_{0}(\xi)d\xi},
\end{align*}
with 
$ b=K_2^2\left\langle \xi_{1}\xi_{2},\left(
-\mathbb{L}\right)  ^{-1}\left(  \xi_{1}\xi_{2}\right)  \right\rangle$. Using that $\Tr(L_1)=1$ we obtain $\displaystyle \frac{\left\langle L_1 \xi G_0,\xi \right\rangle}{\int\xi^{2} G_{0}(\xi)d\xi}=\frac{1}{3}$. Therefore, 
\begin{equation*}
\beta_{t}=\frac{2\beta}{3t}-\frac{8b}{3} \frac{\beta^{\frac{\gamma}{2}+1}}{ \mu(t)}.
\end{equation*}
Using the change of variables $\beta=t^{\frac{2}{3}} y(t)$, we obtain that  $y$ solves
\begin{equation*}
y_{t}=-\frac{8b}{3} \frac{t^{\frac{\gamma}{3}}y^{\frac{\gamma}{2}+1} }{ \mu(t)}.
\end{equation*}
 This differential equation can be solved using elementary methods. Then, after some computations we arrive at 
 \begin{equation*}
 \beta(t)\sim \left(\frac{4\gamma b}{(\gamma+6)}t^2 \right)^{-\frac{2}{\gamma}}, \quad \text{as}\quad t\to\infty.
 \end{equation*}
 Notice that $\frac{\beta(t)^{\frac{\gamma}{2}}}{\mu(t)}\to 0$ as $ t\to\infty$ which validates the consistency of the previous expansion.
 
\bigskip

\bigskip


\section{Table of results and conclusions}
\label{sec:tableconcl}

In \cite{JNV1} and in this paper we have obtained
several examples of long time asymptotics for homoenergetic flows of the
Boltzmann equation. These flows yield a very rich class of possible behaviors.
Homoenergetic flows can be characterized by a matrix $L\left(  t\right)  $
which describes the deformation taking place in the gas. The behavior of the
solutions obtained in this paper depends on the balance between the hyperbolic
terms of the equation, which are proportional to $L\left(  t\right)$, and the
homogeneity of the collision kernel. Roughly speaking the flows can be
classified in three different types, which correspond to the situations in which
the hyperbolic terms are the largest ones as $t\rightarrow\infty$ (cf.~\cite{JNV3}), the case considered in this paper in which the
collision terms are the dominant ones and the case where the collision and the hyperbolic terms have a similar order of
magnitude (cf.~\cite{JNV1}).
\medskip

We summarize here the information obtained about homoenergetic flows.

\begin{itemize}
\item \textbf{Simple shear.}

The critical homogeneity corresponds to $\gamma=0$, i.e. Maxwell molecules.
\bigskip%

\begin{tabular}
[c]{cc}%
\parbox{7 cm}{\hspace{1cm}\underline{Critical case} \vspace{1mm}
\hspace{1.4cm} ($\gamma=0$) } &
\parbox{7cm}{\hspace{0.7cm}\underline{Supercritical case}  \vspace{1mm}
\hspace{1.2cm}($\gamma>0$)}
\end{tabular}

\vspace{4mm}%

\begin{tabular}
[c]{c|c}%
\parbox{7cm}{
Self-similar solutions \vspace{1.3 mm}
with increasing temperature
} & \parbox{7 cm}{
Maxwellian distribution with \vspace{1.3 mm}
time dependent temperature \vspace{1.3mm}
\hspace{1.5mm}
(Hilbert expansion)
}
\end{tabular}


\bigskip

\item \textbf{Homogeneous dilatation.}

The critical homogeneity corresponds to $\gamma=-2$. \bigskip%

\begin{tabular}
[c]{cc}%
\parbox{7cm}{\hspace{1cm}\underline{Critical case} \vspace{1mm}
\hspace{1.4cm} ($\gamma=-2$) } &
\parbox{7cm}{\hspace{0.7cm}\underline{Subcritical case}  \vspace{1mm}
\hspace{1.2cm}($\gamma<-2$)}
\end{tabular}

\vspace{4mm}%

\begin{tabular}
[c]{c|c}%
\parbox{7cm}{
Maxwellian distribution with \vspace{1.3 mm}
time dependent temperature \vspace{1.3mm}
\hspace{1.5mm}
(Hilbert expansion)
} & \parbox{7cm}{
Maxwellian distribution with \vspace{1.3 mm}
time dependent temperature \vspace{1.3mm}
\hspace{1.5mm}
(Hilbert expansion)
}
\end{tabular}
\bigskip


\item \textbf{Planar shear.}

The critical homogeneity corresponds to $\gamma=0$, i.e. Maxwell molecules.
\bigskip%

\begin{tabular}
[c]{cc}%
\parbox{7cm}{\hspace{1cm}\underline{Critical case} \vspace{1mm}
\hspace{1.4cm} ($\gamma=0$) } &
\parbox{7cm}{\hspace{0.7cm}\underline{Subcritical case}  \vspace{1mm}
\hspace{1.2cm}($\gamma<0$)}
\end{tabular}

\vspace{4mm}%

\begin{tabular}
[c]{c|c}%
\parbox{7cm}{ \hspace{1.8mm}
Self-similar solutions \vspace{1.3 mm}
} & \parbox{7cm}{
Maxwellian distribution with \vspace{1.3 mm}
time dependent temperature \vspace{1.3mm}
\hspace{1.5mm}
(Hilbert expansion)
}
\end{tabular}
\bigskip

\item \textbf{Planar shear with $K=0$. 
 }

The critical homogeneity corresponds to $\gamma=0$, i.e. Maxwell molecules.
\bigskip%

\begin{tabular}
[c]{cc}%
\parbox{7cm}{\hspace{1cm}\underline{Critical case} \vspace{1mm}
\hspace{1.4cm} ($\gamma=0$) } &
\parbox{7cm}{\hspace{0.7cm}\underline{Subcritical case}  \vspace{1mm}
\hspace{1.2cm}($\gamma<0$)}
\end{tabular}

\vspace{3mm}%

\begin{tabular}
[c]{c|c}%
\parbox{7cm}{ \hspace{1.8mm}
Self-similar solutions \vspace{1.3 mm}
} & \parbox{7cm}{
Maxwellian distribution with \vspace{1.3 mm}
time dependent temperature \vspace{1.3mm}
\hspace{1.5mm}
(Hilbert expansion)
}
\end{tabular}
\bigskip

\item \textbf{Cylindrical dilatation.}

In this case we have two critical homogeneities: $\gamma=-\frac{3}{2}$ and
$\gamma=-{2}$. \bigskip%

\begin{tabular}
[c]{cc}%
\parbox{7cm}{\hspace{1cm}\hspace{0.8cm} ($\gamma> -{2}$) } &
\parbox{7cm}{\hspace{0.7cm}\hspace{1.2cm}($\gamma<-\frac{3}{2}$)}
\end{tabular}

\vspace{3mm}%

\begin{tabular}
[c]{c|c}%
\parbox{7cm}{ \hspace{1.8mm}
Frozen collisions \vspace{1.3 mm}
} & \parbox{7cm}{
Maxwellian distribution with \vspace{1.3 mm}
time dependent temperature \vspace{1.3mm}
\hspace{1.5mm}
(Hilbert expansion)
}
\end{tabular}
\bigskip

\item \textbf{Combined shear in orthogonal directions $(K_{1}, K_{2}, K_{3})$
with $K_{1}K_{3}\neq0$.}

The critical homogeneity corresponds to $\gamma=0$, i.e. Maxwell molecules.
\bigskip%

\begin{tabular}
[c]{cc}%
\parbox{7cm}{\hspace{1cm}\underline{Critical case} \vspace{1mm}
\hspace{1.4cm} ($\gamma=0$) } &
\parbox{7cm}{\hspace{0.7cm}\underline{Supercritical case}  \vspace{1mm}
\hspace{1.2cm}($\gamma>0$)}
\end{tabular}

\vspace{3mm}%

\begin{tabular}
[c]{c|c}%
\parbox{7cm}{ \hspace{1.3mm}
Non Maxwellian distribution \vspace{1.3 mm}
} & \parbox{7cm}{
Maxwellian distribution with \vspace{1.3 mm}
time dependent temperature \vspace{1.3mm}
\hspace{1.5mm}
(Hilbert expansion)
}
\end{tabular}
\bigskip
\end{itemize}

In \cite{JNV1} we proved rigorously the existence of self-similar solutions
yielding a non Maxwellian distribution of velocities in the case in which the
hyperbolic terms and the collisions balance as $t\rightarrow\infty$. A
distinctive feature of these self-similar profiles is that the corresponding
particle distribution does not satisfy a detailed balance condition.

In this paper we obtained some conjectures about homoenergetic flows in the
case of collision-dominated behavior. When the collision terms are the dominant ones as
$t\rightarrow\infty,$ we have formally obtained that the corresponding
distribution of particle velocities for the associated homoenergetic flows can
be approximated by a family of Maxwellian distributions with a changing
temperature whose rate of change is obtained by means of a Hilbert expansion.

When the hyperbolic terms are much larger than the collision terms the
resulting solutions yield much more complex behaviors than the ones that we
have obtained in the previous cases. This case is discussed in detail in \cite{JNV3}.

We have rigorously proved in  \cite{JNV1} the existence of self-similar solutions in the case in which the collision and the hyperbolic term balance. These solutions, as well as the asymptotic formulas for the solutions obtained in this paper in the case of collision-dominated behavior, give interesting insights about the
thermomechanical properties of Boltzmann gases under shear, expansion or compression
in nonequilibrium situations.  Of particular interest is the sensitive dependence of this behavior on the
atomic forces near the critical homogeneity.  The results of this paper
suggest many interesting mathematical problems which deserve further
investigation. In particular, it would be relevant to determine the precise properties of the collision kernels which allow to prove rigorously the existence of solutions of the Boltzmann equation with the asymptotic properties obtained in this paper and
to understand their stability properties.

Finally we remark that there are also homoenergetic flows 
yielding divergent densities or velocities at some finite time. These flows have not been considered neither in \cite{JNV1}, \cite{JNV3} nor in
this paper but they have interesting properties and their asymptotic behavior would be worthy of study in the future.

\bigskip

\textbf{Acknowledgements. }  We thank Stefan M\"uller, who motivated us to study this problem, for
useful discussions and suggestions on the topic.  
The work of R.D.J. was supported by ONR (N00014-14-1-0714), AFOSR (FA9550-15-1-0207), NSF (DMREF-1629026), and the MURI program (FA9550-18-1-0095, FA9550-16-1-0566). 
A.N. and J.J.L.V. acknowledge support through the
CRC 1060 \textit{The mathematics of emergent effects }of the University of
Bonn that is funded through the German Science Foundation (DFG).\bigskip

\bigskip


\end{document}